\documentclass[manuscript,screen,sigconf]{acmart}
\AtBeginDocument{%
  \providecommand\BibTeX{{%
    \normalfont B\kern-0.5em{\scshape i\kern-0.25em b}\kern-0.8em\TeX}}}

\usepackage{todonotes}
\usepackage{tablefootnote}
\usepackage{multirow}
\usepackage{soul}
\usepackage{xcolor}
\usepackage{lipsum}
\usepackage{xspace}
\usepackage{cleveref}
\usepackage[utf8]{inputenc}
\usepackage{enumitem}
\crefformat{section}{\S#2#1#3} 
\crefformat{subsection}{\S#2#1#3}
\crefformat{subsubsection}{\S#2#1#3}
\usepackage[normalem]{ulem}
\newcommand\red[1]{\textcolor{black}{#1}}
\newcommand\blue[1]{\textcolor{black}{#1}}

\copyrightyear{2025}
\acmYear{2025}
\setcopyright{rightsretained}
\acmConference[Arxiv '25]{}
\acmBooktitle{Arxiv '25}

\begin{document}




\title[\systemname]{Thing2Reality: Transforming 2D Content into Conditioned Multiviews and 3D Gaussian Objects for XR Communication}



\settopmatter{authorsperrow=4}

\author{Erzhen Hu}
\affiliation{%
  \institution{University of Virginia}
  \city{Charlottesville}
  \state{VA}
  \country{USA}
  \postcode{78229}}
  \authornote{Project conducted when the first author interned at Google  in 2024.}
\email{eh2qs@virginia.edu}

\author{Mingyi Li}
\affiliation{%
  \institution{Northeastern University}
  \city{Boston}
  \state{MA}
  \country{USA}
  \postcode{78229}}
\email{li.mingyi2@northeastern.edu}

\author{Jungtaek Hong}
\affiliation{%
  \institution{University of Virginia}
  \city{Charlottesville}
  \state{VA}
  \country{USA}
  \postcode{78229}}
\email{rsv5fd@virginia.edu}

\author{Xun Qian}
\affiliation{\institution{Google Research}
 \city{Mountain View}
 \state{CA}
 \country{USA}}
\email{xunqian@google.com}

\author{Alex Olwal}
\affiliation{%
  \institution{Google Research}
  \city{Mountain View}
  \state{CA}
 \country{USA}
  }
\email{olwal@acm.org}

\author{David Kim}
\affiliation{%
  \institution{Google Research}
  \city{Zurich}
  \country{Switzerland}
  }
\email{kidavid@google.com}

\author{Seongkook Heo}
\affiliation{%
  \institution{University of Virginia}
  \city{Charlottesville}
  \state{VA}
  \country{USA}
  \postcode{78229}
  }
\email{seongkook@virginia.edu}

\author{Ruofei Du}
\authornote{Corresponding author.}
\affiliation{%
  \institution{Google Research}
  \city{San Francisco}
    \state{CA}
  \country{USA}
  }
\email{me@duruofei.com}




\newcommand{\erzhen}[1]{\todo[inline,color=blue!20!white]{\textbf{EZ:} #1}}


\definecolor{darkgreen}{rgb}{0,0.5,0}
\definecolor{orange}{rgb}{1,0.5,0}
\definecolor{teal}{rgb}{0,0.5,0.5}
\definecolor{darkpurple}{rgb}{0.5, 0, 0.5}
\definecolor{lightergrey}{RGB}{225,225,225}

\newcommand {\xun}[1]{{\color{purple}\bf{$\leftarrow$ XQ: #1}\normalfont}}
\newcommand {\ruofei}[1]{{\color{orange}\bf{$\leftarrow$ RD: #1}\normalfont}}
\newcommand {\yinda}[1]{{\color{teal}\bf{$\leftarrow$ YZ: #1}\normalfont}}
\newcommand {\david}[1]{{\color{darkpurple}\bf{$\leftarrow$ DK: #1}\normalfont}}
\newcommand {\alex}[1]{{\color{darkgreen}\bf{$\leftarrow$ AO: #1}\normalfont}}
\newcommand{\karthik}[1]{{\color{cyan}\bf{$\leftarrow$ KR: #1}\normalfont}}
\newcommand{\mingyi}[1]{\textbf{\textcolor{darkgreen}{ML: #1}}}

\newcommand{\textbt}[1]{\textbf{\textit{#1}}}
\newcommand{\tocite}[1]{{\color{purple}[CITE: #1]\normalfont}}
\newcommand {\systemname}{Thing2Reality\xspace}
\newcommand {\conceptname}{Continuum of Dimensions\xspace}

\newcommand{\etal}{et al.\xspace}
\newcommand{\ie}{\emph{i.e.}\xspace}
\newcommand{\eg}{\emph{e.g.}\xspace}
\newcommand{\vs}{\emph{v.s.}\xspace}

\newcommand {\bt}[1]{\textbf{#1} \normalfont}

\newcommand{\squishlist}{
 \begin{list}{$\bullet$}
  { \setlength{\itemsep}{0pt}
    \setlength{\parsep}{3pt}
    \setlength{\topsep}{3pt}
    \setlength{\partopsep}{0pt}
    \setlength{\leftmargin}{1.5em}
    \setlength{\labelwidth}{1em}
    \setlength{\labelsep}{0.5em} 
  }
}

\newcommand{\squishend}{\end{list}}

\newcommand\greybox[1]{%
  \vskip\baselineskip%
  \par\noindent\colorbox{lightergrey}{%
    \begin{minipage}{0.99\columnwidth}#1\end{minipage}%
  }%
  \vskip 0.5em%
}

\newcommand{\bbox}[1]{
  \vskip\baselineskip%
  \noindent\fbox{\begin{minipage}{0.95\columnwidth}#1\end{minipage}}
  \vskip 0.5em%
}

\newif \ifhighlight \highlighttrue    

\providecolor{added}{RGB}{65,105,225}   
\providecolor{deleted}{RGB}{255,36,0}     

\newcommand{\added}[1]{{\ifhighlight {{\color{added}{}#1}} \else {#1}\fi}}
\newcommand{\deleted}[1]{{\ifhighlight {{\color{deleted}\sout{#1}}}\fi}}

\begin{abstract}
During remote communication, participants often share both digital and physical content, such as product designs, digital assets, and environments, to enhance mutual understanding.
Recent advances in augmented communication have facilitated users to swiftly create and share digital 2D copies of physical objects from video feeds into a shared space.
However, conventional 2D representations of digital objects restricts users' ability to spatially reference items in a shared immersive environment.
To address this, we propose Thing2Reality, an Extended Reality (XR) communication platform that enhances spontaneous discussions of both digital and physical items during remote sessions.
With Thing2Reality, users can quickly materialize ideas or physical objects in immersive environments and share them as conditioned multiview renderings or 3D Gaussians. 
Thing2Reality enables users to interact with remote objects or discuss concepts in a collaborative manner.
Our user study revealed that the ability to interact with and manipulate 3D representations of objects significantly enhances the efficiency of discussions, with the potential to augment discussion of 2D artifacts.
\end{abstract}

\ccsdesc[500]{Human-centered computing~Collaborative and social computing}

\keywords{extended reality, augmented communication, image-to-3D, information artifacts, multi-modal, remote collaboration}

\begin{teaserfigure}
    \centering
  \includegraphics[width=\columnwidth]{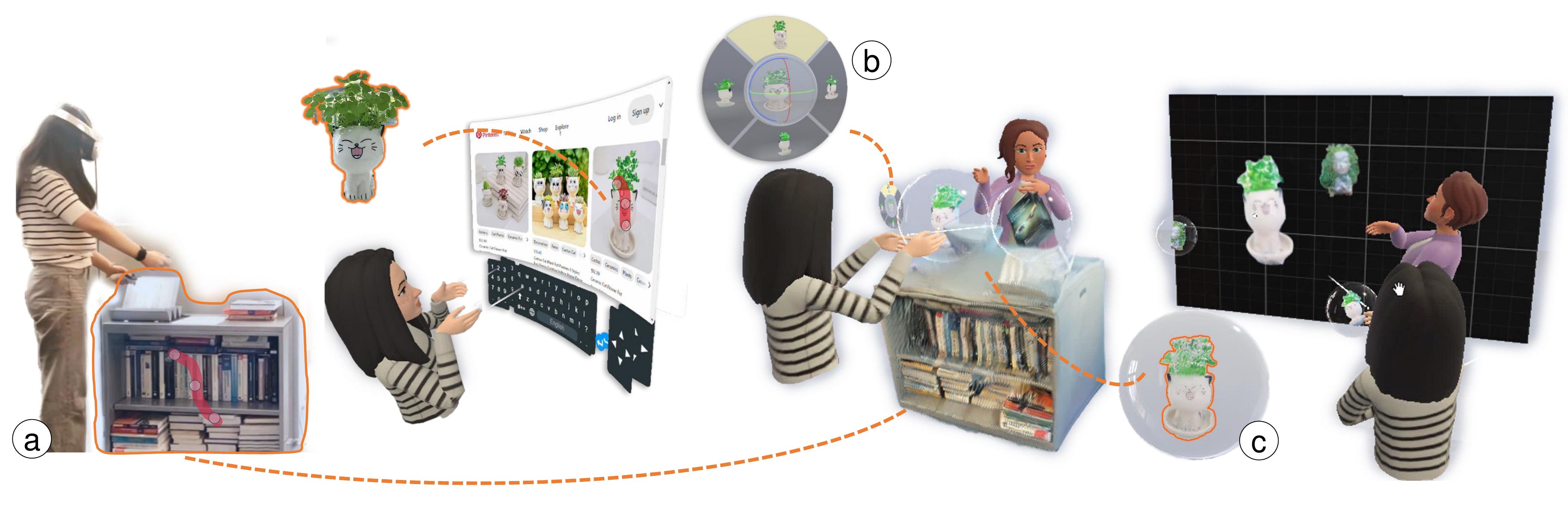}
    \caption{An example use case of Thing2Reality. Alice and Charlie are discussing room decorations in a shared XR space (b). Alice begins by bringing a shelf from her physical office (a) into the virtual environment. She then searches for a cute cat planter using the web browser interface. With Thing2Reality, she summons 3D Gaussian of the planter and places it onto the virtual shelf. Alice and Charlie then engage in a discussion about various planter designs, projecting 3D Gaussian representations of the planters (c) onto a whiteboard in the space. This allows them to transform 2D images into interactive 3D objects, which can be collectively viewed, manipulated, and compared in real-time, facilitating a seamless and collaborative ideation process.
    }
    \label{fig:teaser}
\end{teaserfigure}

\maketitle

\section{Introduction}
Shared artifacts, including digital resources (\eg, text, images, videos), and physical objects (\eg, prototypes, printouts), play a crucial role in facilitating effective communication and idea generation.
They provide common spatial reference points that bridge gaps between collaborators, enhancing creative exploration and ideation~\cite{brereton2000observational}.
Besides physical artifacts, designers frequently use online platforms like Pinterest and Google to source relevant digital artifacts that can support their design processes~\cite{herring2009getting}. 
However, using shared artifacts in remote meetings often pose challenges, especially in scenarios that require quick and spontaneous sharing, such as brainstorming sessions. 
First, artifacts shared via remote meetings are typically in 2D, whether they are captured via camera or retrieved from online repositories, limiting the understanding compared to interactions with physical objects or 3D models.
Second, in physical meetings, participants can easily observe and interact with tangible artifacts, which facilitates creative exploration and idea generation processes \cite{brereton2000observational}. However, in remote meetings, this level of interaction is often unavailable or limited. 

Several methods have attempted to address these challenges, such as preparing 3D models in advance via CAD or 3D scanning \cite{izadi2011kinectfusion}, or employing specialized real-time 3D capture setups \cite{orts2016holoportation,10.1145/3332165.3347872}. 
While effective, these approaches have limitations: pre-made 3D assets do not support \textit{spontaneous} sharing, and specialized setups are often impractical for general use.
Recent advances in AI-driven text-to-3D and image-to-3D technologies~\cite{tang2024lgm} present an accessible and efficient alternative, lowering barriers to 3D content creation and enabling broader participation in collaborative efforts.

In this paper, we aim to understand how image-to-3D AI can mediate users in XR communication and how the design of such a system can benefit users and integrate interactive image-to-3D workflows within various Extended Reality (XR) meeting context. We introduce Thing2Reality, an XR communication platform that enables fluid interactions with 2D and 3D artifacts. Thing2Reality allows users to segment content from any source (video streams, shared digital screens) within the XR environment (\autoref{fig:teaser}a), generate multi-view renderings (\autoref{fig:teaser}b), and transform them into shared 3D Gaussians for interactive manipulation (\autoref{fig:teaser}c).

We evaluated Thing2Reality in a user study involving three tasks: avatar decoration, furniture arrangement, and workspace organization. Our findings suggest that while 3D objects facilitate intuitive explanations and hands-on collaboration, 2D representations are more often used in final pitch deliverables, suggesting a trade-off between the two formats depending on the context and purpose of the task.
We demonstrate applications of Thing2Reality in various workspace and social scenarios, highlighting how on-the-fly 3D generation can enrich interaction and social connectedness, augmenting human-human communication regarding shared digital and physical artifacts. 

In summary, we contribute: 
\begin{itemize}
    \item \textbf{Thing2Reality, an XR communication system} that provides on-the-fly AI-mediated 3D objects generation by enabling users to present and share spontaneous thoughts, and augment their shared digital and physical artifacts with remote peers. 
    \item \textbf{Thing2Reality applications}, which offer insights for future XR collaborative interface design.
    \item \textbf{Findings from a comparative user study} ($N$=12) evaluating the effectiveness of Thing2Reality in supporting spontaneous 2D-to-3D object generation, compared to 2D snapshots from digital and physical sources.
    \item \textbf{Findings from an exploratory user study} ($N$=18) examining the use of Thing2Reality (both 2D-to-3D and 3D-to-2D workflow) for discussing and presenting both 2D and 3D objects in XR communication.
\end{itemize}

\section{Related Work}
Our work is inspired by prior art on vision-language interfaces, distributed communication, and task-space collaboration. 

\subsection{Vision-Language Interfaces}
\subsubsection{Generative Models}

Text-to-3D and image-to-3D methods, such as DreamFusion~\cite{poole2022dreamfusion}, focus on score distillation sampling (SDS) that utilizes pretrained 2D diffusion models to generate 3D content, but faces problems with speed and diversity. Recent advances in large reconstruction models~\cite{li2023instant3d,hong2023lrm} use non-SDS methods. Large Gaussian Models (LGM)~\cite{tang2024lgm} use similar methods to ~\cite{kerbl20233d}, with algorithms to convert 3D Gaussian into meshes. 
Advances using multi-view diffusion models as a prior have also made generation of complex, textured 3D models possible. These generative models provide a foundation for transforming 2D visuals into 3D representations.

\subsubsection{Text-Based and Spatial-Oriented Prompting}
Prompting has been enabled by large language models primarily as a natural language-oriented way of interaction. However, text-based prompts are constrained by the level of control they can provide, especially in terms of the spatial aspects of images~\cite{li2023multimodal}. Modern prompting in computer vision and machine learning has evolved towards spatial-oriented prompting. For example, ControlNet~\cite{zhang2023adding} enables users to condition image generation using additional images such as depth maps and human skeletons. Segment-Anything (SAM)~\cite{kirillov2023segany} enables the use of points and boxes in addition to text as a way of prompting mask segmentation.

To enable users to grab content from shared multimedia artifacts and transform them into 3D generated objects, Thing2Reality utilizes a controllable pipeline that allows users to explicitly identify areas of interest using spatial-oriented prompts, such as points and strokes.

\subsubsection{Vision and Language in Computer Interfaces}
Recent work has explored vision-language methods, such as text-to-image generation~\cite{liu20233dall}, in 3D design workflows and image generation for news illustration~\cite{liu2022opal}, and in mediating human-human co-creation with creativity support tools~\cite{10.1145/3411764.3445219, deshpande2020towards, ghosh2019interactive}. 
For communication-related areas, Visual Captions~\cite{liu2023visual} utilized language input to retrieve relevant images as visual aids to augment human-human communication. These images can be used as shared media between users to facilitate communication. However, the existing visuals of shared media and artifacts have not been fully exploited in these interfaces.

In this work, we used both text prompting and spatial-oriented prompting for identifying objects of interest and sought to understand how AI-mediated 3D artifacts can contribute to XR communication and collaboration.

\subsection{Distributed Communication and Task Space Collaboration}
People increasingly use remote conferencing platforms~\cite{10.1145/3491102.3517558,10.1145/3544548.3581013,10.1145/3544548.3581148,10.1145/3500868.3559470} for workplace meetings, education, entertainment, and social interaction with families and friends.
The field of computer-mediated cooperative work has investigated the importance of shared media~\cite{marlow2016beyond,10.1145/3064663.3064722,10322229,10.1145/3173574.3173855} during in-person and remote communications, especially in understanding how people use, create, and share multimedia artifacts. Shared task spaces are essential for scenarios such as education~\cite{10.1145/3411764.3445323,10.1145/3491102.3517486,10.1145/3411764.3445323}, creativity support~\cite{10.1145/3122986.3123002}, tabletop and tablet games~\cite{10.1145/3626475,10.1145/3623509.3635255, yuan2024field}, video editing~\cite{10.1145/3126594.3126659}, and physical task demonstration~\cite{10.1145/3491102.3501927,10.1145/3332165.3347872}.

\subsubsection{AI-Augmented 2D Shared Task Space}
IllumiShare~\cite{10.1145/2207676.2208333} enabled users to share physical and digital objects on arbitrary surfaces. Recent work such as ThingShare~\cite{10.1145/3544548.3581148} enabled the digital copies of physical objects with deep neural network to mediate remote communication and collaboration. Visual Captions~\cite{liu2023visual} also supported shared media by augmenting language as visual aids between users in remote conferencing. 
Other work has explored mobile sharing~\cite{10.1145/2675133.2675176} and reconstruction of mobile phone video streams~\cite{vanukuru2023dualstream}, but providing a stable view can be challenging.

The advances in generative AI have enabled new opportunities for blending reality and enabling coarse-grained and fine-grained customization of environments. For example, BlendScape~\cite{rajaram2024blendscape} enabled a blended virtual environment by meaningfully merging people's virtual backgrounds using in-painting and image-to-image techniques with selective regeneration of portions in the 2D video-conferencing environment. 
However, prior work has either focused on 2D visual aids with images retrieved by language to augment communication or 2D and 3D reconstruction/blending of capturing physical shared task spaces for remote collaboration. The potential of text-to-3D and image-to-3D workflows to transform generic artifacts into interactive 3D objects for communication tasks remains under-explored.

\subsubsection{Enabling Spatiality in Remote and Video Conferences}
2.5D and 3D video-conferencing has been focused primarily on reconstructing and displaying of life-sized talking heads~\cite{10.1145/3588037.3595385}, gaze-aware 3D photos~\cite{10.1145/3472749.3474785}, and space-aware scene rendering for avatar placement (\eg, VirtualCube~\cite{zhang2022virtualcube}, ChatDirector ~\cite{qian2024chatdirector}). 

Different from 2.5D and 3D avatars, SharedNeRF~\cite{mose2024} leverages photo-realistic and view-dependent
rendering with NeRF and point clouds, which enables an on-the-fly volumetric task space.
Spatiality in video conferencing~\cite{hauber2006spatiality} has been explored by combining person videos with immersive desktop collaborative virtual environments. These spatial interfaces were found to positively influence social presence and co-presence compared to 2D, while potentially compromising task focus and efficiency. Sousa \etal ~\cite{sousa2019negative} explored ways to mediate ambiguity in workspace awareness when interacting with 3D digital assets and found an increase in mental demand when converting coordinates between frames. However, the creation, interaction, and sharing of AI-generated 3D objects to mediate human-human communication in remote settings remain under-explored.

Furthermore, prior work has used extended reality to assist remote physical task guidance using virtual replicas~\cite{10.1145/2807442.2807497,tian2023using}, which was found to be more efficient than 3D annotations in mixed-reality collaboration scenarios~\cite{tian2023using}. However, these studies focused more on the manipulation and interaction of pre-designed 3D virtual objects and the support of virtual replicas for spatial referencing rather than how the spontaneity of 3D object creation can enable better delivery of human thoughts and ideas.

\section{Design Considerations}

\subsection{Design Space}

We present three dimensions in articulating our design space (\autoref{fig:my_design space}) and situate Thing2Reality into prior literature of distributed communication
and demonstration of 2D and/or 3D artifacts.
Prior work also explores different ways of creating pre-made or catalog assets, such as using gestures to approximate and imitate the object~\cite{holz2011data} among a database of known objects, or understanding the role of virtual replicas in communication and remote assistance~\cite{10.1145/2807442.2807497,tian2023using}.
We did not include this line of work because we focused on the spontaneity of sharing things during the communication phase. 

\subsubsection{Methods: Capturing versus Generating}
The distinctions between capturing and generating methods are represented in \autoref{fig:my_design space}: Methods).
\paragraph{\textbf{Capturing as Virtual Replicas}}
This line of  work supports spontaneity of 2D and 3D artifacts sharing via snapshot-based, 3D reconstruction, or search-based methods, which aims to capture and reconstruct the physical reality as virtual replicas~\cite{mose2024,10.1145/3544548.3581148,Lindlbauer2018Remixed,10.1145/3631418}.
For example, ThingShare ~\cite{10.1145/3544548.3581148} explores snapshot-based interactions to facilitate object-focused collaboration.
Some other work utilized real-time 3D reconstruction with sophisticated camera setups  (\eg, ~\cite{Lindlbauer2018Remixed})
or single camera with NeRF (\eg, ~\cite{mose2024}). 
However, this line of research focuses on cloning the physical space for physical tasks or remote assistance, whereas shared information artifacts such as digital images and videos in the web were not explored.
\paragraph{\textbf{Generating as On-the-Fly Assets}}
Different from capturing or reconstructing surrounding scenes or objects as replicas, GenAI enabled new opportunities for remote human-human communication, specifically the task space communication in both digital and physical space.
For example, digital images, and sketches~
\cite{liu2023visual} can be used for augmenting spontaneous communication with LLM-enabled digital search. BlendScape~\cite{rajaram2024blendscape} used stable diffusion and in-painting techniques to blend virtual backgrounds of users together as a meaningful cohabited space.
Different from these methods that enabled 2D visual aids or 2D virtual background, Thing2Reality enables both 2D and 3D with (primarily) image-to-3D methods. 
Furthermore, we separate these artifacts as digital vs. physical for the data source - as these artifacts can be either searched, or drawn in the digital information space (as visual aids like ~\cite{liu2023visual}), or image streams captured directly from the physical environment (for physical tasks or object-focused collaboration). 
\subsubsection{Representation}
\begin{figure}[!htb]

     \centering
     \includegraphics[width=0.9\linewidth]{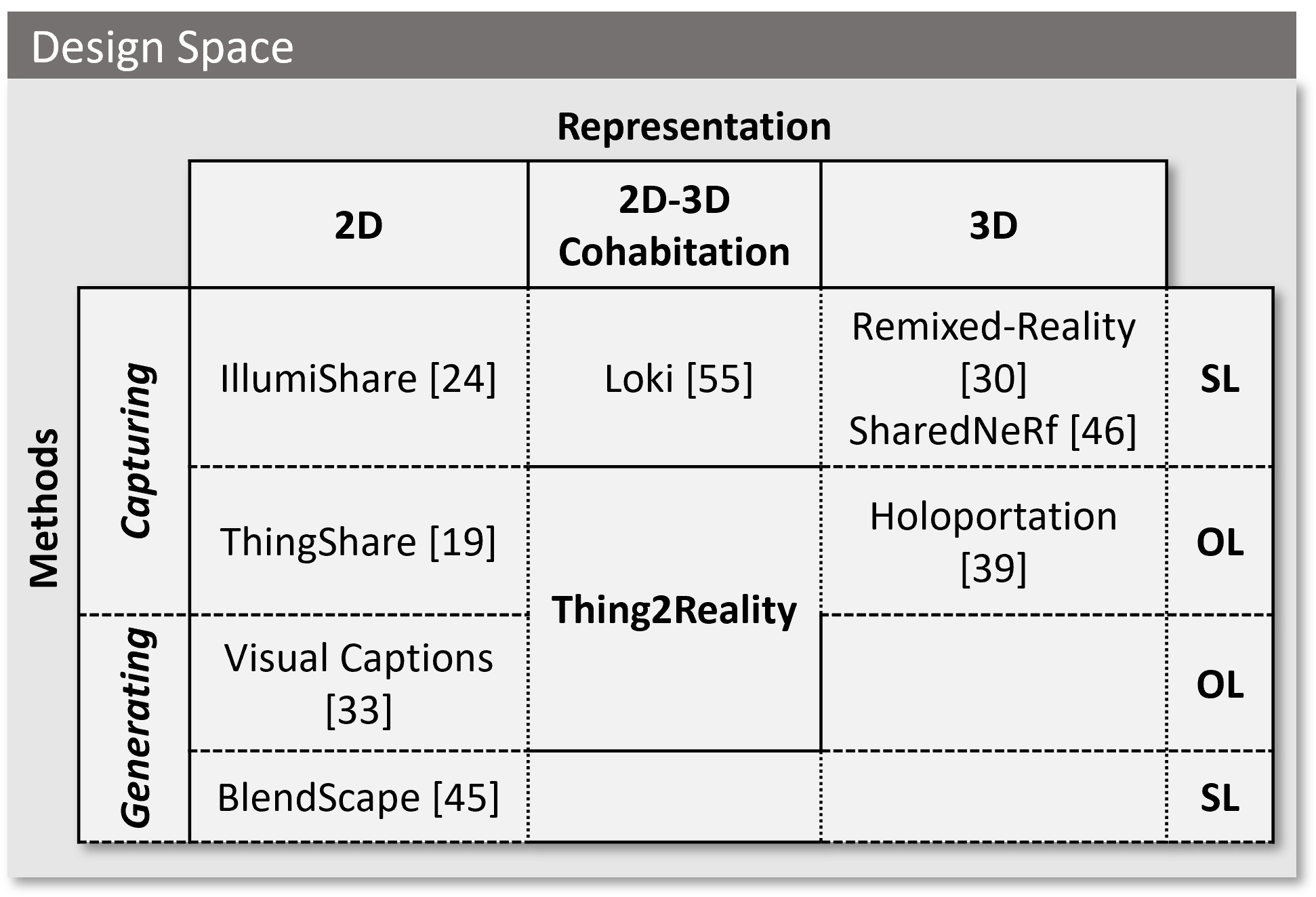}
     \caption{Design Space of Thing2Reality: with the rows of Scene-Level (SL) \& Object-Level (OL) capturing, Scene-Level (SL) \& Object-Level (OL) generating, and columns for representations of objects. Left rows of the table indicate the difference between capturing and generating. Right rows of the table indicate the difference between scene-level and object-level.
     }\label{fig:my_design space}
\end{figure}

The column (\autoref{fig:my_design space}: Representation) differentiates the object representations (2D-only, 2D+3D, 3D-only) supported by the system. 
However, most objects explored by prior individuals or collaborative work and remote conferencing are represented as 2D~\cite{10.1145/3544548.3581148, liu2023visual,10.1145/2207676.2208333}.  Remixed-reality supports real-time 3D reconstruction and provided scene modifications~\cite{Lindlbauer2018Remixed}. Some recent work support 3D scene sharing during remote meetings ~\cite{mose2024}.

The bidirectional interactions between 2D images and 3D models have been explored in augmented reality ~\cite{10.1145/3313831.3376233}, 
yet the 3D models are pre-loaded rather than spontaneously identified by the user, and are thus not included in the design space. 
Furthermore, this line of work either explores pre-made bidirectional 2D and 3D object transformation from either the physical environment ~\cite{10.1145/3379337.3415849,10.1145/3472749.3474769} or from digital assets~\cite{10.1145/3313831.3376233,10.1145/3242587.3242597}. Furthermore, most of them emphasized individual editing rather than collaborative interactions. An exception is Loki~\cite{10.1145/3332165.3347872} that provides different data modalities (both 2D videos and 3D point-cloud scenes) of physical spaces for scene-level (SL) interactions with point clouds and videos, whereas Thing2Reality focuses on object-level (OL) interactions.

\begin{figure}
     \centering
     \includegraphics[width=\linewidth]{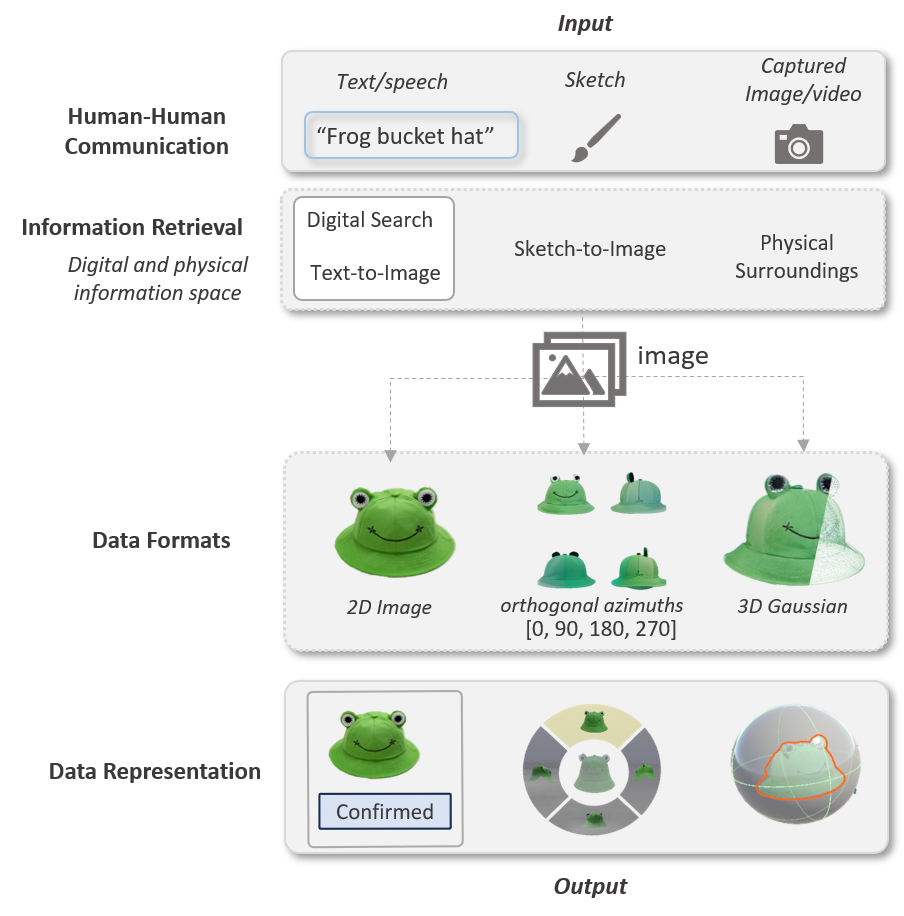}
     \caption{
     Human-human communication methods: 1) text or speech, 2) sketch, 3) images or videos can be used as \textit{input} to achieve ideal 2D images via digital search, image/video capturing, or GenAI/ML models (\textit{text-to-image}, \textit{sketch-to-image}), which can then be converted to arbitrary segmented image, conditioned multiview renderings, and 3D Gaussian. 
     }
     \label{fig:my_workflow_illustrate}
\end{figure}

In brief, we emphasized the role of spontaneity in object-level generation and cohabitation of 2D and 3D artifacts by helping users identify objects of interests from both digital information space and physical space with generative methods.

\subsection{Design Goal}
Based on the design space, we formulated the following three objectives to direct the design of Thing2Reality to enable efficient and flexible discussion around artifacts in XR communication.

\paragraph{
\textbf{DG1 Spontaneity: Enable Spontaneous Communication Using Digital and Physical 3D Artifacts As Visual Aids}} Acknowledging the importance of both physical and digital artifacts in professional discussions, we aim to facilitate a seamless conversion of 2D artifacts from diverse data sources into 3D representations. These inputs from a variety of sources includes digital files (such as images, and sketches) and physical objects captured via camera feeds.

\paragraph{\textbf{DG2 Cohabitation: Support for Co-Habitation of 2D and 3D Objects During Communication.}}
Conventional remote conferencing approaches often rely on a single modality of capture and presentation data (\eg, 2D images, 2D videos) to teach or guide remote participants~\cite{10.1145/2501988.2502045}. Loki~\cite{10.1145/3332165.3347872} demonstrates the potential benefit of incorporating multiple data modalities. To facilitate efficient discussions and collaborative sessions, Thing2Reality should allow for multiple data representations.
By supporting interaction with multiple data modalities, users can choose the most appropriate representation for their current communication needs, leading to more effective collaboration.

\paragraph{\textbf{DG3 Transition: Enable Flexible Bi-Directional \red{Transformations} Among Digital Media Forms (\ie, 2D images, videos, and 3D)}} 
XR workspaces can be more dynamic and open compared to traditional videoconferencing, and the presence of multiple data modalities may introduce friction for users. Recognizing the diverse needs of remote collaboration with multiple data modalities in \textbf{D2}, it is also essential to allow users to frictionlessly switch between different forms of these representations (2D images, videos, multi-view representations, and 3D models) according to the context of their discussion. This 
requirement would imply that Thing2Reality should not only store and organize various forms of media but also allow for their easy retrieval and transformation during discussions. By enabling flexible bi-directional transitions between digital media forms (2D-to-3D, 3D-to-2D), users can adapt their communication style to the specific requirements of the task at hand, leading to more efficient and effective collaboration.
\section{Thing2Reality System Overview}

A key takeaway from our design space highlighted the effectiveness of integrating 3D object affordances with the spatial organization advantages of 2D artifacts. The workspace encompasses not just flat artifacts but also three-dimensional things. We developed Thing2Reality to capitalize on the strengths of 3D artifacts to facilitate communication between individuals, focusing on the utilization of surrounding information surfaces (e.g., tables, whiteboards). This system introduces the capability to seamlessly transition 3D artifacts from 2D digital or physical counterparts. Before delving into the system's design, it is crucial to clarify the definition of ``Thing'' in the context of our work.

\subsection{What the ``Thing''? Exploring the Role of User-Generated 3D Assets in Spontaneous Communication}
In bringing user-generated 3D assets into distributed human-human communication, bridging 2D and 3D counterparts may shape a new way of communication in the immersive information space. We aim to outline the typical approaches individuals take when spontaneously incorporating various artifacts (\ie, sketches, searched images, and physical objects) into discussions as a source of inspiration, explanation, or clarification (\autoref{fig:my_workflow_illustrate}).

\begin{figure*}
    \centering
    \includegraphics[width=\linewidth]{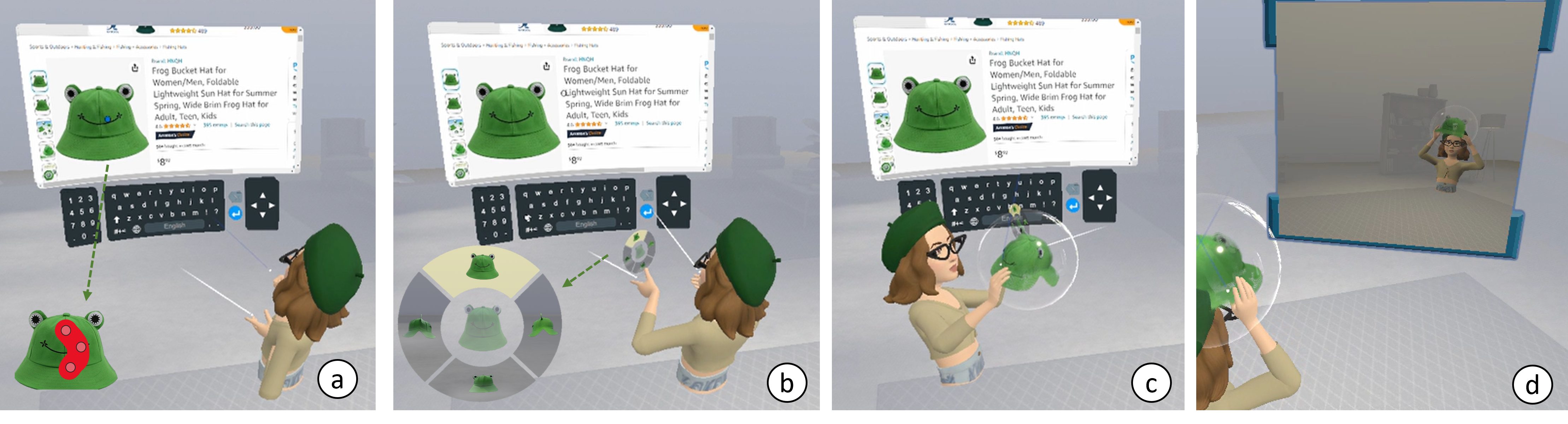}
    \caption{An example user journey: (a) a user begins by selecting preferred visuals to bring to reality. This is achieved by painting on the desired region within the web browser or camera feed of the physical space. These objects are subsequently processed through progressive stages: starting from a 2D segmented image, evolving into conditioned multi-view renderings, and ultimately, to a 3D Gaussian representation. (b) Meanwhile, orthogonal views are laid out along the rings of the \textit{Pie Menu}.
    (c) The 3D Gaussians are summoned after 1-2 seconds. (d) The user can re-position and re-scale it via the \textit{Sphere Proxy}.
    }
    \label{fig:workflow}
\end{figure*}

\begin{itemize}[leftmargin=*]
    \item \textbf{Text-based content (\autoref{fig:workflow} - 1)}: Text-based content uses words and language to convey ideas, including written descriptions, transcribed speech, and notes. Text-based contents can be transformed into 2D images use text-to-image methods like Gemini Imagen. 
    \item \textbf{Hand-created visual content (\autoref{fig:workflow} - 2):} Hand-created visual content encompasses manually produced images, diagrams, or visual representations, either physical or digital. This includes sketches, drawings, and hand-drawn diagrams, providing intuitive and spontaneous representations of ideas, spatial relationships, or abstract concepts in communication. Current methods such as ControlNet~\cite{zhang2023adding} use sketches as one of the ways for controlling image generation.
   \item \textbf{Digital visual content (\autoref{fig:workflow} - 1):} Images found through online searches like Google images or Pinterest, screenshots, and digital artwork stock photos to find images that closely align with their discussion topics, utilizing these images as a reference point~\cite{herring2009getting}. 
    \item \textbf{Captured real-world content (\autoref{fig:workflow} - 3)}: Photographs or scans of physical objects and environments, which can serve as a powerful means of conveying ideas, but their integration poses challenges for distributed users~\cite{10.1145/3544548.3581148,brereton2000observational}, who might opt to digitally capture and share these items. \red{It is important to note that these digital 3D representations of real-world content do not always capture the specific details of an object as accurately as a virtual replica (\eg, NeRF).  Instead, they serve as a proxy for the original object. Furthermore, this can be beneficial for items when part of an object's side is not easily capturable, or when it's difficult to photograph at close range (e.g., a large shelf).}
\end{itemize}

Transforming these variations of 2D content into 3D objects can help enhance the immediacy and tangible engagement with abstract concepts, such that users can gain a deeper mutual understanding during discussions.
\red{Furthermore, text-to-image and sketch-to-image generation methods often produce less predictable results due to the vagueness of the input in describing expected images, making precise control challenging. In contrast, searching for existing images or capturing real-world content allows for more direct selection and accuracy. This difference in control stems from the interpretative nature of AI-based generation versus the specificity of human-curated or directly captured visual content. Recognizing this difference in control between AI-based generation and human-curated or directly captured visual content, Thing2Reality's design primarily focuses on \textbf{digital visual content} and \textbf{captured real-world content} with image-to-3D approaches. This ensures more precise control over the images used in communication, enhancing the system's reliability and user experience. }

\paragraph{\textbf{Our Focus on Spontaneous Human-Human Communication in XR:}} 
In light of these insights from our design space and prior literature, our research focuses on the exploration of user-generated 3D assets. This focus is driven by the unique potential of these assets to disseminate the essence of abstract concepts to distributed XR users.

We aim to explore the \textit{spontaneity of user-generated 3D objects} from any sources in facilitating distributed XR communication.

\subsection{Interaction Workflow}
Here we show the default interaction workflow with the example of digital search. 


\begin{figure}[!htb]

     \centering
     \includegraphics[width=\linewidth]{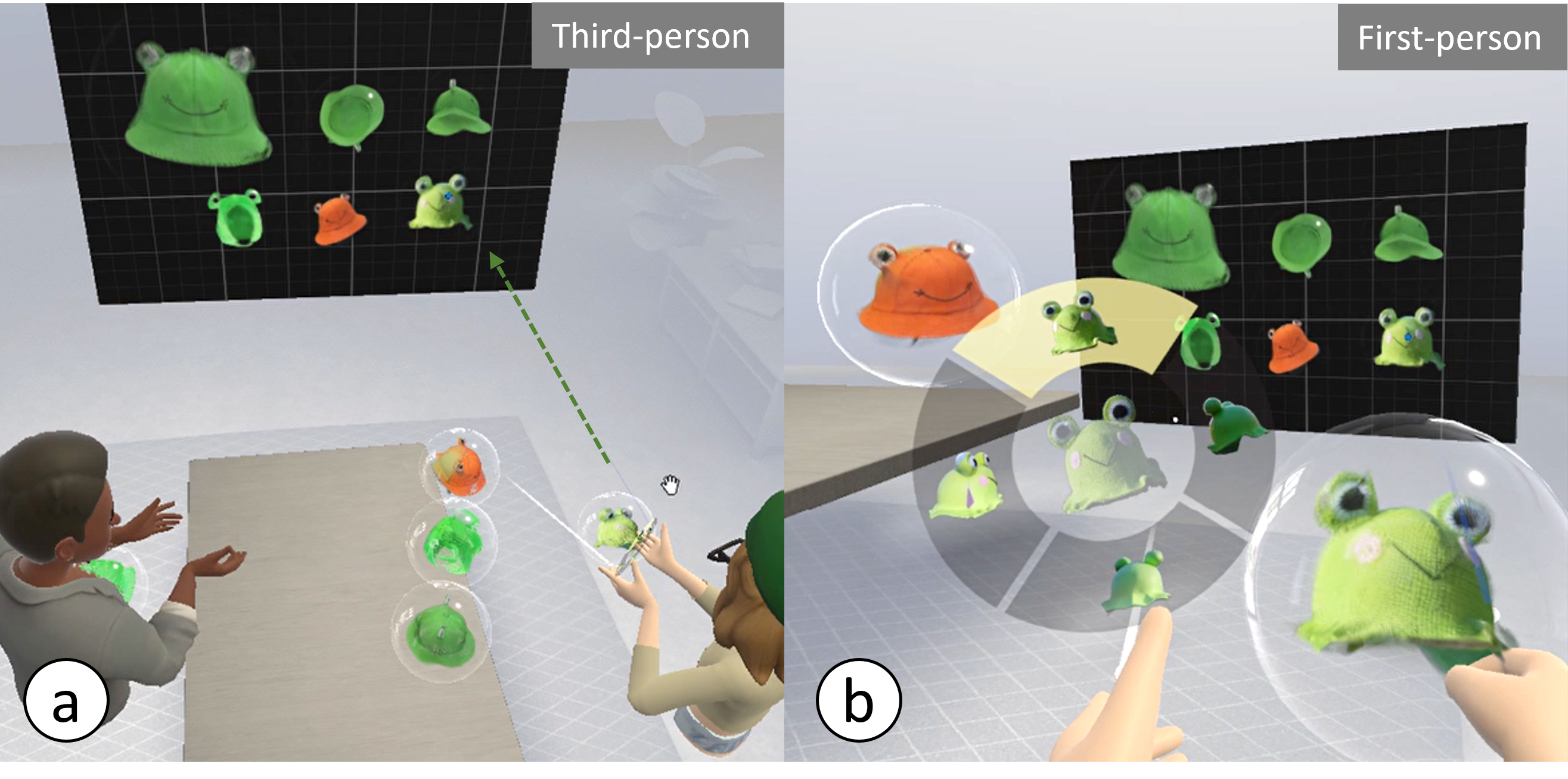}
     \caption{3D-to-2D Process: A user can capture snapshots from different perspectives of the 3D Gaussians, and project it on the whiteboard. (a) Third-person perspective; (b) first-person perspective.
     }\label{fig:workspace}
\end{figure}
\begin{figure}
     \centering
     \includegraphics[width=\linewidth]{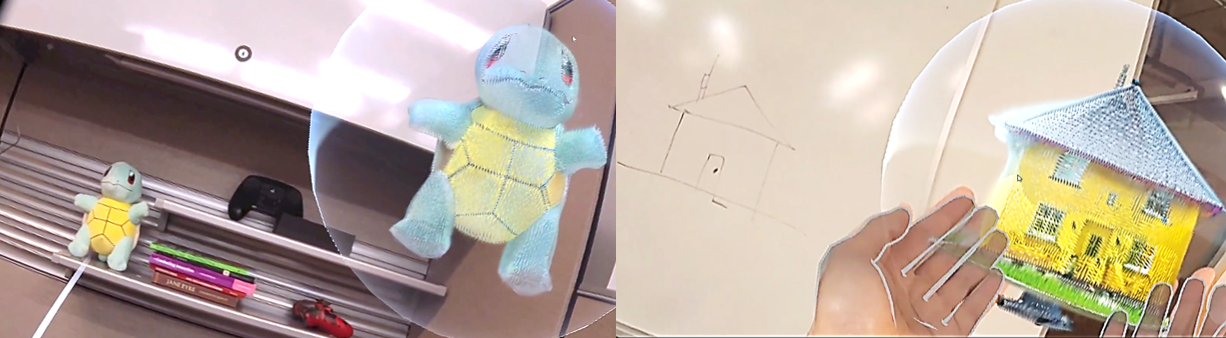}
     \caption{With video see-through mode, users can bring physical objects and sketches to the shared space in XR.}
    \label{fig:Augmented}
\end{figure}

\paragraph{\textbf{Interactive Object Segmentation}}
The user can identify object of interest by holding the grip button of the controller while capturing a pointdown and point up event with the trigger button, which results in three point positions identified on the data source image (\eg, a web browser view, or a camera feed). The object of interest will then be segmented from the image source, and provide the user the segmented results besides their hands. The user can then confirm to send the images for multi-view rendering and 3D Gaussian creation. 

\paragraph{\textbf{2D-to-3D: Creation of Multi-views and 3D Generated Objects}}
After the user confirms the object of interest,
the multiple conditioned views will be rendered on a 2D \textbf{Pie Menu} (\autoref{fig:workflow}b) attached to the user's left controller. The center of the Pie Menu shows the original image being cropped from the data source (web-views, images from physical space) or generated from the image generation model.
The four orthogonal views, generated with conditioned diffusion models, will be displayed on the top (front view), left (side views), right (side views) and bottom (back views) of the outer ring of the Pie Menu. Selecting the central image also displays a 360° video of the object.
The user can show or hide it by pressing the ``X'' button on the controller. 
The generated 3D object will then become a shared object in the environment, which can be moved, grabbed, and re-scaled by all the users via the semi-transparent \textbf{Sphere Proxy}  (\autoref{fig:workflow}c) as a collider.
The orthogonal views of a 3D object on the 2D Pie Menu are private to the user who created it, but it can be converted from any shared 3D objects generated in the environment. 

\paragraph{\textbf{3D-to-2D: Projecting Things to Surrounding Whiteboard and Table for Workspace Communication}}

The user can take a snapshot from any angles of the generated 3D objects, under the field of view of the user, and project the point of view on collaborative surfaces like a whiteboard or a table (\autoref{fig:workspace}), which is different from the \textit{discrete} orthogonal views due to the \textit{continuous} perspectives a 3D object presents. 

Users can use raycasting to drag things around the whiteboard, rescaling things by selecting the object and using the y axis of the thumbstick to make it larger or smaller. They can also delete the object by selecting it and then pressing on the ``B'' button on the controller. 
Users can also select the discrete orthogonal views on their private 2D menu, which can be projected on the whiteboard. The central image can be projected on the whiteboard to show 360-degree videos of the object. 

\section{Implementation}
\label{sec:implementation}

\begin{figure*}
    \centering
    \includegraphics[width=1.05\linewidth]{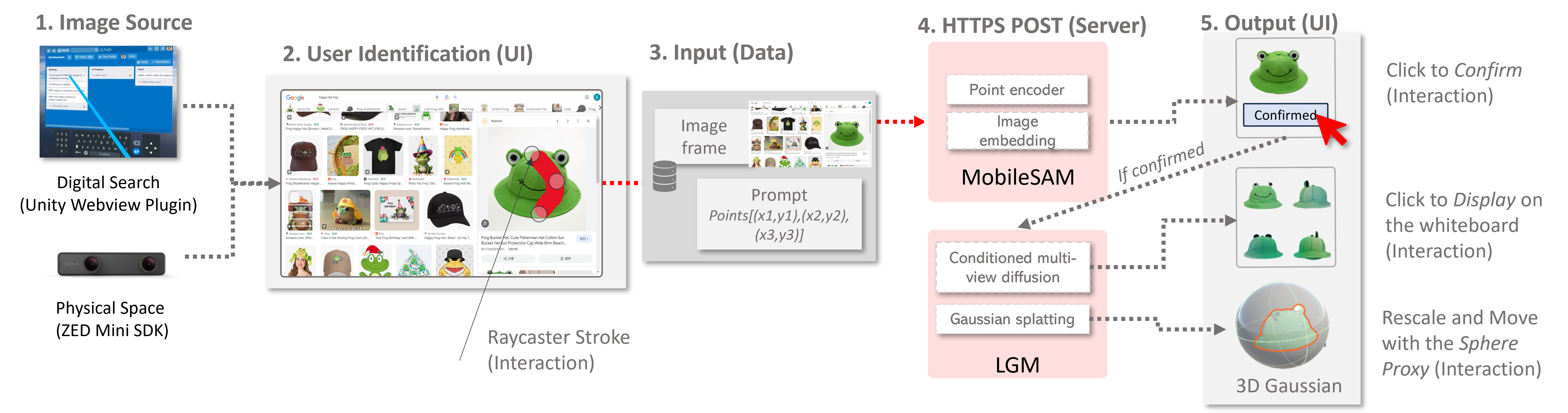}
    \caption{Implementation Diagram of Object-Level Thing-to-3D between Unity and Python Flask Server. 
    }
    \label{fig:Diagram}
\end{figure*}

The virtual environment was developed using Unity 2022.3.19f1 and the following SDKs: Oculus Interaction Toolkit, Meta Avatar SDK for rendering avatars, gestures, and lip-syncing, and Photon Fusion and Voice SDK for voice streaming between avatars, and ZED SDK for physical space sensing.

\paragraph{\textbf{System Setup}}
The study program ran on a desktop PC with an Intel
Core i7-13700K processor and an NVIDIA RTX 4070 Ti GPU for MobileSAM~\cite{zhang2023faster}, text-conditioned~\cite{shi2023mvdream} and image-conditioned~\cite{wang2023imagedream} multi-view diffusion models, and Large Gaussian Models~\cite{tang2024lgm} to fuse multiview renderings into interactive 3D Gaussians.   

Due to the privacy issues of capturing data from physical environments via current passthrough technologies of VR/MR HMD, a ZED Mini camera was attached in front of the Meta Quest 3.

\paragraph{\textbf{3D Object Rendering Pipeline}}

The enhancement of extended reality communication involves a blend of digital and physical elements, captured through both web-based sources and external camera feeds. 

Utilizing a webview feature within a virtual space allows for the seamless incorporation of digital content into 3D artifacts. This is achieved by implementing a Unity web browser plugin~\cite{vuplex}. To merge physical reality into the XR experience, image frames are captured using the ZED Mini camera\footnote{ZED Mini: \url{https://store.stereolabs.com/products/zed-mini}}. These frames are then integrated into the Unity environment, providing a real-world context that users can interact with alongside digital content.

Hence, users can identify objects in both web-based content and live camera feeds capturing physical spaces via ZED and turn the identified 2D images into 3D. This includes the ability to perform actions such as making strokes or taking snapshots. Interactions within the virtual environment are managed through a custom event listener via a Python Flask server. The original input (\autoref{fig:Diagram}:3) is the selected image frame with the three points on the selected image frame filtered from the user's stroke interaction using the raycaster of the Meta Quest 3 controller.

The Unity application communicates with the Python Flask Server via HTTP POST requests (\autoref{fig:Diagram}:4), activating models for quick, gesture-prompted segmentation. This process identifies objects of interest based on user interactions.
Once segmented, objects are visualized within the Unity environment for user confirmation. Multi-view rendering and 3D Gaussian modeling techniques are employed to create more interactive representations of these objects. The 3D Gaussian output was a ``.ply'' format, but can also be represented as a 2D video, or a visual effect in the Unity environment. 
The 3D Gaussian were then imported and visualized in Unity as Gaussian splats, surrounded by a semi-transparent sphere around it as a proxy collider to enable interactions like Grab, Scale, and Move with hands or controllers.

\section{Application Scenarios}

Thing2Reality can be used in a diverse set of applications to elevate the experiences of the human-human communication across workspace and social gatherings. 

\subsection{Workspace Discussion}
\subsubsection{Collaborative Concept Explanation}

Thing2Reality enhances concept explanation by providing tangible 3D explorations of ideas that may be difficult to convey with static 2D images. For example, in a product pitch meeting (\autoref{fig:scenarios}a), a designer could quickly generate a 3D model from a sketch or reference image. This allows stakeholders to interact with the concept from various angles, facilitating more informed feedback and effective brainstorming.
\subsubsection{3D Design Discussion and Co-Creativity}
In early prototyping stages, Thing2Reality can transform 2D references into 3D models, enhancing design discussions. For instance, an interior designer could create 3D furniture models or room layouts from reference images (\autoref{fig:teaser}). This enables clients to better visualize proposed designs and collaborate more effectively on decisions about materials, colors, and spatial arrangements, fostering co-creativity between designer and client.

\subsection{Customization for XR Avatars, Virtual Try-ons, and 3D Emojis}

\subsubsection{Avatar Decoration and Virtual Try-ons}

In XR meetings, personalizing avatars and environments is crucial for immersive experiences. Traditional GUI-based customization can be limiting. Thing2Reality enables users to quickly create or customize 3D objects for avatars to wear or interact with (\autoref{fig:scenarios}c), such as hats, bags, or coffee mugs. This fosters a more natural and engaging environment for various social gatherings in XR spaces such as coffee chats, family meetings, and gatherings around friends.

\subsubsection{Personalized 3D Emojis and Memes}
\begin{figure}[!htb]

     \centering
     \includegraphics[width=\linewidth]{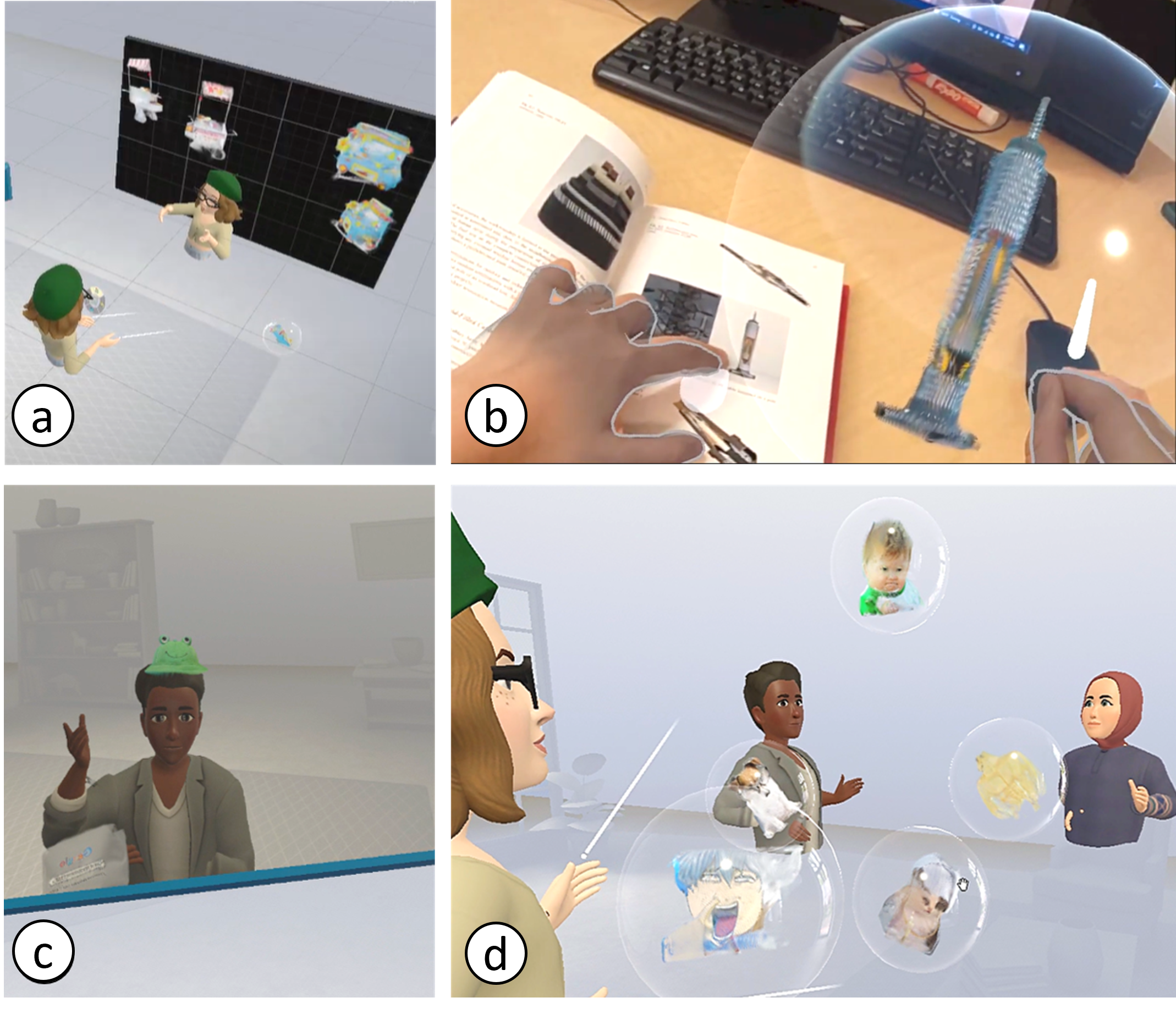}
     \caption{Application scenarios: a) Workspace discussion; b) Magic book for communication c)
     Avatar decoration, and d) 3D Memes for subculture.
    }
    
    \label{fig:scenarios}
\end{figure}

While 2D emojis, GIFs, and memes are widely used in online communication, Thing2Reality allows users to transform online images or memes into the 3D version (\autoref{fig:scenarios}d). This feature enhances social gatherings, and VR live streams, moving beyond pre-designed 3D emojis like those in Meta Workroom.
Different from Apple's newest Genmojis\footnote{Genmojis: \url{https://www.apple.com/newsroom/2024/06/introducing-apple-intelligence-for-iphone-ipad-and-mac/}} that enabled 2D text-to-emoji in text-based digital interactions, Thing2Reality enabled more intuitive expressions during vr meetings. 

\subsubsection{Magic Book} Thing2Reality can be used to create a MagicBook effect~\cite{10.1145/634067.634087}, augmenting book contents to aid discussions and enhance storytelling. For example, in a children's book about dinosaurs, Thing2Reality could bring illustrations to life as interactive 3D models, allowing young readers to explore the creatures' anatomy and interact with them in a virtual environment. This approach creates a more engaging and immersive reading experience for long distance social  play~\cite{yuan2024field,10.1145/3544548.3580720}.

\begin{figure}
    \centering
    \includegraphics[width=0.9\linewidth]{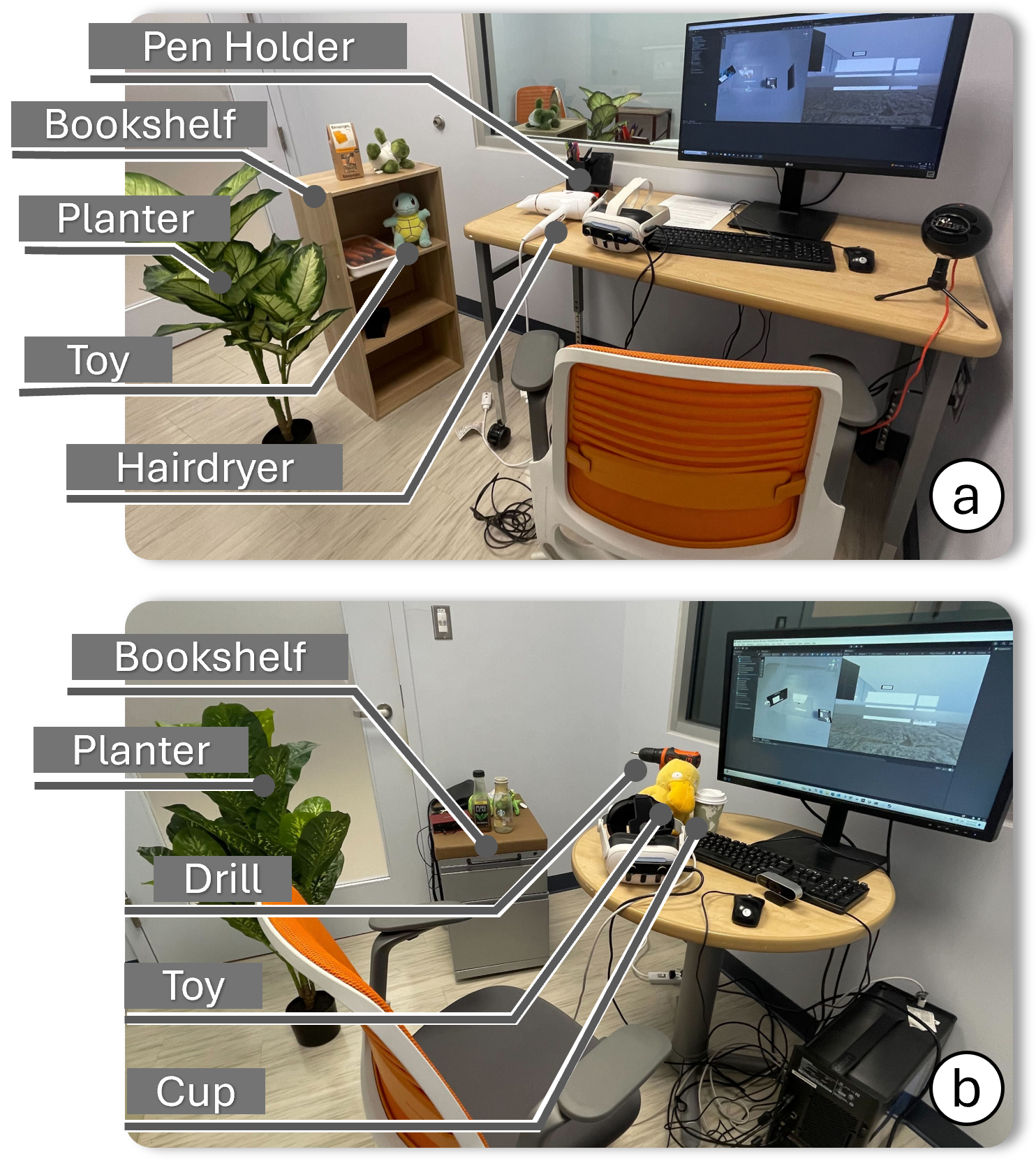}
    \caption{Study Setup for the two rooms (a-b), where each room includes a planter, a bookshelf/cabinet, a electronic object (hairdryer/drill), a toy.}
    \label{fig:roomsetup}
  
\end{figure}
    

\section{Comparative User Study}
\label{sec:controlled}
We conducted a comparative user study to understand how Thing2Reality improves communication around the shared information artifacts (both physical and digital items), compared to 2D objects.

\subsection{Study Design}
The study employed a within-subject design that compared 3D (Thing2Reality) and 2D (Baseline) shared artifacts during impromptu discussions. Both conditions allowed for manipulation of objects (2D vs. 3D) upon confirmation of 2D interactive segmentation from data input sources. Two data sources were tested in all tasks: capturing things from a private screen with digital search, and from the video camera feed \textbf{(DG1)}.
The design of the 2D condition was inspired by ThingShare~\cite{10.1145/3544548.3581148}, with a difference that the \textit{2D shared segmented artifacts} could be manipulated in the 3D space using VR controllers. In the 3D condition, users can manipulate \textit{3D shared Gaussian objects} with VR controllers. The study was IRB-approved.
\begin{quote}
    \textbf{RQ1:} How do the differences between the generated 3D objects and 2D virtual replica affect users' understanding, exploration, and trust of an object when both can be similarly manipulated in a VR environment?

    \textbf{RQ2:} How do creating and using 3D Gaussian objects from digital and physical sources affect the effectiveness of discussions about the objects? 
    
\end{quote}
For dependent variables, we used a questionnaire to assess users' experiences with Thing2Reality in shared task spaces. Key measures included satisfaction with the format, comfort with object control, trust in object representation, ease of communication, and effectiveness in conveying and understanding object details. We evaluated these factors for both sharing (as the presenter) and understanding (as the inquirer) object information. The questionnaire also addressed the system's impact on clarifying complex concepts and improving overall subject understanding. Two authors collaboratively analyzed the qualitative data using Affinity Diagramming to identify main themes in users' responses and statistically analyzed the survey data using a paired-t test between 2D and 3D condition ($\alpha=0.05$). 
\begin{figure*}
    \centering
    \includegraphics[width=1\linewidth]{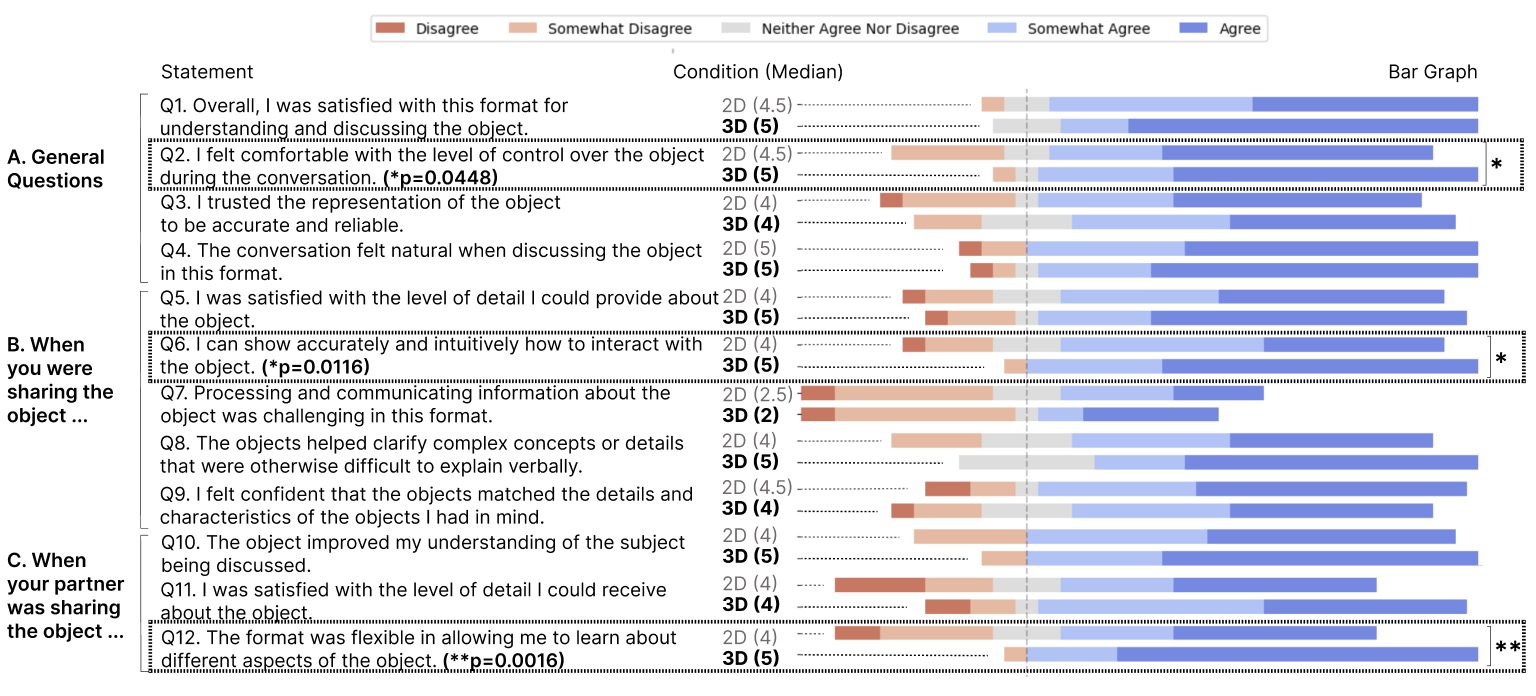}
    \caption{Summary of Survey Results. Statements were divided into three sections: A) general, B) presenter role, and C) inquirer role.}
    \label{fig:survey1}
\end{figure*}

\subsection{Participants}
We recruited 12 participants (4 female, 8 male) as pairs from the university via email lists. Participants reported medium familiarity with VR (Median=3, IQR=1; scale 1-5). Their VR/AR experiences primarily involved gaming ($N=4$), online browsing ($N=2$), and research participation ($N=2$). 
Two participants noted prior social meetup experience in AR/VR platforms.
The in-person study took around 90 minutes per session. Each participant was compensated with \$20 USD.

\subsection{Study Setup and Procedure}
Pairs of participants were invited to two rooms (see \autoref{fig:roomsetup}) so that they could not see each other nor the physical objects in the other room. All the sessions were video-recorded.

After completing each condition and context, participants responded to an intermediate survey about the drawbacks and benefits of the tried 2D or 3D format and the effectiveness of the system in completing the tasks using a 5-point Likert scale and open-ended questions. After completing all tasks, a final survey collected user preferences over different conditions and subjective feedback about the features.
\paragraph{\textbf{Tutorial and WalkThrough (30 min)}}
The experimenter first demonstrated how to use the system. After explaining the VR environment and basic interactions, the experimenter demonstrated how a user can create 2D and 3D virtual objects from an image on a web browser, and from the physical space captured through the ZED Mini camera mounted on the VR headset. Participants were then asked to replicate the demonstration themselves to become familiar with the controls.
\paragraph{\textbf{Task 1: Showcase and Inquiry Task (30 min)}}
The aim of this task is to understand how 3D objects compare to 2D images in facilitating communication, object manipulation, and information sharing during spontaneous searches and discussions.
This task simulates a brainstorming session where a Presenter spontaneously develops an idea for an abstract concept or object, searches for it online, and shares it with a distributed Inquirer. 
The Inquirer was instructed to ask the Presenter three specific questions: 1) Could you show me all the components of the object? 2) 
Could you show me how you would like to interact with the object?
3) an open-ended question.
After completing these inquiries, the participants switch roles.
In the 2D condition, the first Presenter captures a toy from the real world and selects an animal through digital search. The second Presenter finds a cooking tool digitally and a drill from the real world.
For the 3D condition, the first Presenter scans a hairdryer from the real world and chooses any cooking tool digitally. The second Presenter captures an animal digitally and finds a toy from the real world.

\paragraph{\textbf{Task 2: Collaborative Floor Planning Task: (30 min)}}
The goal of this collaborative task is to explore how participants interact and make decisions when arranging both physical and digital objects in a simulated living space followed by prior work~\cite{10.1145/3173574.3173620}. This study compared 2D and 3D conditions.
Participants work in pairs to decorate a living room using a mini 2D floor map. The task involves placing a mix of real-world items (such as planters and toys) and digitally sourced furniture on the map.
In the 2D condition, participants bring three items: a digitally searched chair, a real-world bookshelf (subject to debate), and a digital object of their choice to place on the bookshelf (placement). After arranging these on the mini-floor plan, participants engage in a brief debate over which bookshelf to keep.
The 3D condition follows a similar structure but uses different items. Each participant bring a real-world planter, a digitally searched tea table (for debate), and a real-world object to place on the tea table (placement). 

 \begin{figure*}
    \centering
    \includegraphics[width=0.9\linewidth]{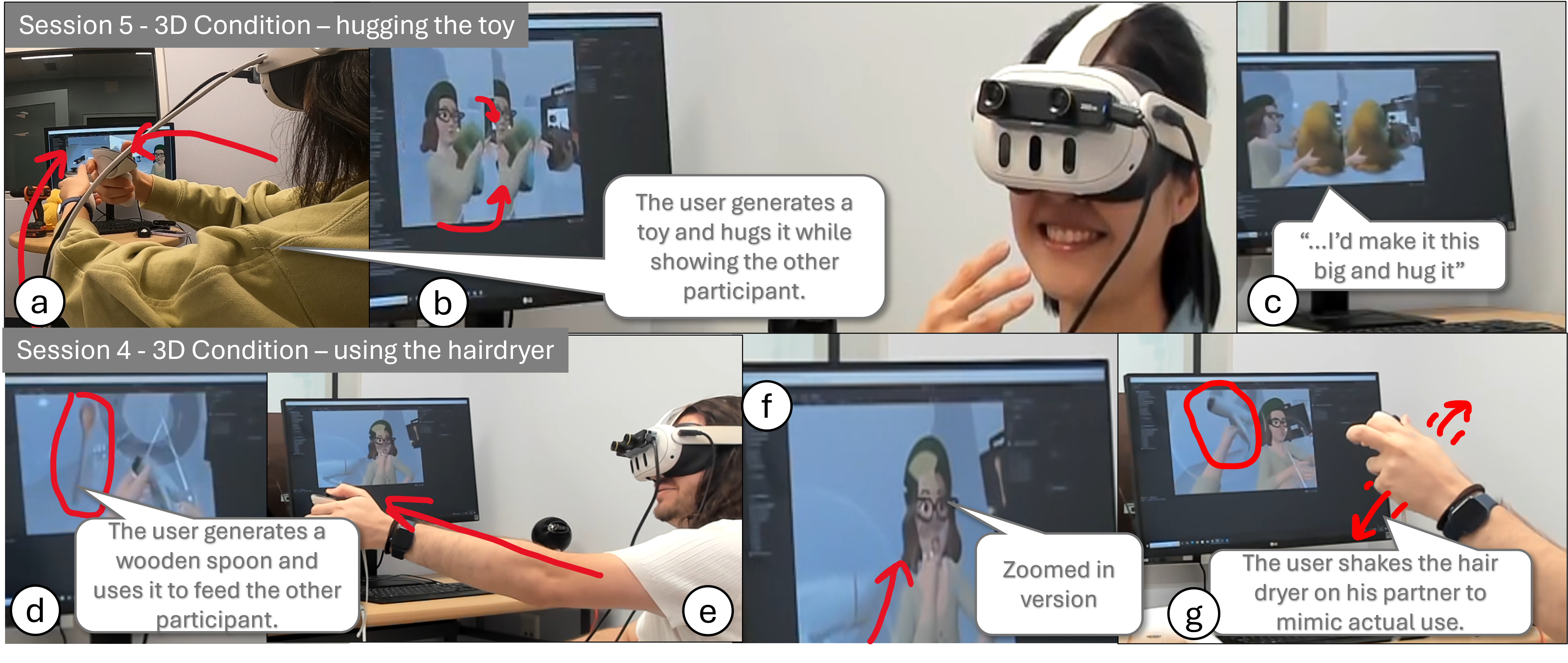}
    \caption{This sequence shows how the use of 3D generated objects allowed users to interact and demonstrate objects intuitively; and how the use of 3D generated objects enabled a-b) (In one deployment) showcasing hugging actions with 3D Gaussian objects from the real world to the virtual space; c) enlarging the object; d-g) another deployment: showcasing the use of real life objects (a hairdryer).}
    \label{fig:action}
\end{figure*}

\subsection{Findings}
\subsubsection{Questionnaire Data (\autoref{fig:survey1})}
Overall, participants were satisfied with using both 2D and 3D formats to understand and discuss the objects with their partners, with the 3D format having a slight but not significant advantage ($Q1: 3D: Median=5, IQR=1$) over 2D format ($Q1: 2D: Median=4.5, IQR=1$). 
Participants reported that the 2D/3D object representation was accurate and reliable ($Q3: 2D: Median=4, IQR=2.5; 3D: Median=4, IQR=2$), exploring 2D/3D objects was simple and intuitive, and discussing them was also a natural process ($Q4: 2D: Median=5, IQR=1; 3D: Median=5, IQR=1$). In particular, there is a significant difference in participants' sense of control over 2D/3D objects, where participants felt that they had higher level of control over 3D objects ($Q2: Median=5, IQR=1, p<0.05$)  than 2D objects ($Median=4, IQR=2$). 

When participants acted as Presenters, they found it easy and effective to describe both 2D and 3D objects, with no significant challenge in communication ($Q7: 2D: Median=2.5, IQR=2; 3D: Median=2, IQR=3.5$), and were able to clarify object-related information ($Q8: 2D: Median=4, IQR=2$; $3D: Median=5, IQR=1.5$). Although not statistically significant, participants found 3D objects more effective for description and communication, and interactions were shown more accurately with 3D ($Q6: 2D: Median=4, IQR=2; 3D: Median=5, IQR=1; 
p<0.05$). While participants were satisfied with the level of detail in both formats ($Q5: 2D: Median=4, IQR=2; 3D: Median=5, IQR=1$), they felt 2D objects matched expected details better ($Q9: 2D: Median=4.5, IQR=1$) than 3D objects ($Q9: 3D: Median=4, IQR=2$).
Here, we present findings on communication dynamics, expectations regarding 2D versus 3D objects, and the comparative effectiveness of digital versus physical objects.

When participants acted as Inquirer, they reported that both 2D and 3D objects improved their understanding of the subject (Q10; 2D: Median=4, IQR=1; 3D: Median=5, IQR=1). Participants were satisfied with the level of details (Q11; 2D: Median=4, IQR=3; 3D: Median=4, IQR=1) and were able to grasp key features quickly. The 3D representation showed a slight but not significant advantage on such object understanding. 
In particular, 3D objects ($Q12: 3D: Median=5, IQR=0; p<0.005$) were significantly more flexible in supporting participants learning about different object aspects  than 2D objects ($Q12: 2D: Median=4, IQR=3$). 
This could be attributed to the comprehensive 3D view and flexible object manipulation, such as rotation.

\subsubsection{General Feedback}
Overall, in the Showcase and Inquiry task, when participants were Presenter, 10 out of 12 participants preferred using 3D format over 2D to communicate with their partners and when they were Inquirer, 9 participants preferred 3D format. In the Collaborative Floor Planning task, 10 participants preferred interacting with 3D objects.
\paragraph{\textbf{Facilitating Effective Communication via Contextual Visualization and Shared Perspectives}}
In both 3D and 2D settings, participants commented that the generated object helped make conversations more engaged and effective by establishing in-context visualization (3D) and creating shared views (2D).
Participants found that 3D objects helped simulate the real-life scenario by providing depth understanding and preview of how objects would be situated in space. For example, in the Floor Planning Task, participants were able to quickly have a sense of the furniture arrangement and its spatial configuration. P11 described it as an intuitive process: ``\textit{I can quickly have a sense of how the room's gonna look like when we put the furniture there, very intuitive}''. 

\blue{Furthermore, our observations revealed that users naturally employed 3D objects to showcase actions in the virtual environment.  For instance, the sequence (\autoref{fig:action}) illustrates the functionality of Thing2Reality, and some of its uses by participants. For a showcase task with 3D condition, P9 captured a toy duck from her physical environment to showcase how she would interact with it to her partner (P10). 
P9 first performed a hugging gesture with the virtual duck (\autoref{fig:action}a), which was shown from the P10's perspective (\autoref{fig:action}b). P9 then enlarged the toy duck to hug a giant version of it, saying that she "\textit{... would make [the toy] this big and hug it}'' if it were possible to do so in real life (\autoref{fig:action}c). During another deployment, P7 found a wooden spoon (\autoref{fig:action}d) from Google Search. He then demonstrated a feeding action directed towards P8 by moving his hand towards the P8's mouth (\autoref{fig:action}e-f). He started another interaction, miming the use of a hair dryer on the P8. He waved the virtual hair dryer with side-to-side motion typical of real-life hair drying (\autoref{fig:action}g). } 
Different from deployments of 3D conditions that showcased actions, participants mainly pointed to the 2D images when holding them in their hand.

In particular, 3D objects were significantly more flexible in supporting participants learning about different object aspects ($3D: Median=5, IQR=0; p<0.005**$) than 2D objects ($2D: Median=4, IQR=3$) (Q12). 
The 3D nature of the objects proved particularly beneficial, offering a comprehensive view that covered all aspects of the item with continuous rotations. This feature enhanced the experience for both presenters and inquirers, \eg, P2 explained, ``\textit{I could turn the object around and look at the different features of it and show how I can interact with it better.}'' 
For participants acting the Inquirer role, the 3D representation helped in overall understanding and resolved potential confusions. P9 noted, ``\textit{I could also understand better when the presenter was pointing at some part of the object.}'' This suggests that the 3D format facilitated clearer communication and more effective object exploration between participants that
``\textit{mimics the real world...}'' (P8).

Despite the constraints of depth and volume for 2D objects, participants liked the shared view of 2D with less dimensions that helped participants present and understand ideas.
As a Presenter, P2 stated that ``\textit{Looking at the picture and having the person see it too made it easier to be sure they have a good idea what I am presenting about}'' (P2). 
During inquiring, most participants also felt that the 2D format was easy to understand and helped them quickly get an idea of the object, yet most of them concerned about that they cannot see it from all the viewpoints during the presentation.

\paragraph{\textbf{2D Object Replicas vs. 3D Object Generations with User Expectations}}
Although participants found 3D generated objects realistic and helped mutual understanding, they recognized the weakness that the generated objects missed some features that they expected, or contained inaccurate or blurred details, making the result misaligned with their expectations. 
This is also aligned with the survey findings where participants reported that 2D objects contained more expected and relevant details~$(Median=4.5, IQR=1)$ than 3D objects~$(Median=4, IQR=2)$ (Q9).

P9 articulated this challenge - ``\textit{The generated part on the back from the digital scanning and the side part of the actual item was not very clear and it had a totally different color from what I expected, which made it a little difficult for me to explain that it originally shouldn't be this way and had to tell what is different from the my expectations}'' (P9). 
This disparity stemmed from the 3D Gaussian model's attempt to predict unscanned sides of physical objects or complete partial digital images, sometimes resulting in unexpected additions or omissions of important features.

In contrast, 2D object representations demonstrated greater effectiveness in retaining crucial details from both physical objects and digital images. This resulted in representations that are aligned with the original items. P3 noted that ``\textit{The object resembled the actual objects more closely and did not lose important details}'' (P3). A few participants mentioned that the preservation of these details in 2D format facilitated clearer communication, \eg, ``\textit{it was simple for me and her to communicate our objects and details with the help of objects in the scene}'' (P1).

These findings suggest that while 3D Gaussian objects offer the advantage of 3D visualization, they face challenges in meeting user expectations for accuracy and detail preservation. In contrast, 2D representations, though lacking in dimensionality, prove more effective in maintaining object fidelity and supporting clear communication.

\paragraph{\textbf{3D Gaussians for 2D-to-3D Generation: Digital vs. Physical Effectiveness}}

Participants reported varying experiences when converting captured digital and physical items into 3D Gaussian representations.
\red{In general, they appreciated the spontaneity that they can turn both physical and digital objects into 3D Gaussian artifacts as visual aids, which is aligned with our design goals (\textbf{DG1}).}
However, the effectiveness of the resulting 3D Gaussian objects on supporting communication differed notably between digital and physical items, with participants highlighting distinct considerations for each approach.

Turning digital searched items into 3D offered participants the flexibility of online search, unrestricted by their physical surroundings.  In these instances, participants were less concerned about unexpected results, as they could not easily find alternative perspectives of the 2D object, where generated viewpoints of 3D Gaussian objects were perceived highly useful. 
This was particularly useful in the Floor Planning Task, where participants could compare online objects with existing items in their space. 
As P1 noted - ``\textit{it's helpful to bring in any objects from amazon for example and see where i want it}'' (P1). However, some participants complained about the unpredictable size ratio, either the output made the online object looked much bigger or it got smaller than they thought.

In contrast, the process of converting physical items into digital 3D Gaussian representations, while limited to available real-life items, resulted in higher user expectations for detail retention and accuracy compared to the digital items.
Participants appreciated its ability to generate 2D/3D outputs from real-life objects, especially for items that is hard to move, or large item that is impossible to capture some viewpoints at a distance. However, they could be disappointed if important details were missing or the quality was low, \eg,
P5 stated that ``\textit{the details of the objects getting omitted}'' (P5).
Despite these issues,  participants valued the ability to use physical objects as references and to check how digital items might fit with their existing possessions.
It also supported participants to use physical objects as a reference and check whether other digital items fit it. P9 emphasized this advantage - ``\textit{I like that I could mix the physical object with the digital objects since I can consider what could go along with the actual stuff that I already own by placing digital items next to the actual objects}'' (P9).

\section{Exploratory User Study}
\label{sec: study-2}

To gain insights about the communication dynamics and user behaviors afforded by Thing2Reality, we conducted an IRB-approved study with another nine pairs of users ($N=18$). 
\red{Different from our comparative study in understanding the pros and cons of separating 3D and 2D (2D versus 2D-to-3D), this study focused on how the coexistence of 2D and 3D artifacts supports communication dynamics. We used tasks like avatar decoration, furniture arrangement, and workspace demonstration, which require 3D artifacts for task completion, were designed to invoke these dynamics.
}

\begin{quote}
    \red{\textbf{RQ3} How do users use 2D and 3D artifacts for spatial arrangements, object-centered discussions, and presentation tasks?}
\end{quote}


Furthermore, the study aimed to evaluate the entire workflow of Thing2Reality's 2D-to-3D and 3D-to-2D processes, as well as to understand users' comprehension, mental efforts, conversational experience, and communication and interaction around 2D and 3D artifacts in an XR environment. 
\red{Since we focused on understanding the communication dynamics and user behaviors afforded by Thing2Reality, where users will use both 2D and 3D artifacts (\textbf{DG2}) and freely transition (\textbf{DG3}) between them (2D-to-3D and 3D-to-2D) for communication and discussion in the virtual environment, regardless of the input (either be digital or captured physical content). We used ``digital search'' as an example for the input method to maintain the consistent fidelity for input methods. 
}

\begin{figure*}
   \centering
     \includegraphics[width=\linewidth]{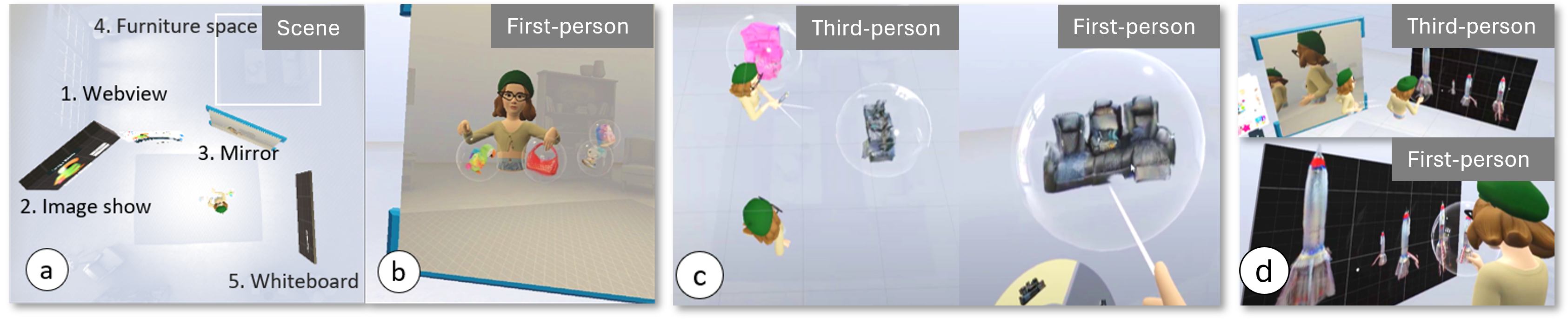}
      \caption{Study Scene and Tasks of Thing2Reality from first-person and third-person views: a) Overview of the immersive environment; b) Warm-up task: avatar decoration; c) Furniture arrangement task; d) Workspace task: Toy pitch. 
      }
    \label{fig:Tasks}
    \end{figure*}

\subsection{Participants}
We recruited 18 participants (7 female, 11 male) aged between 22-29 ($\bar{x}=26.1$) through the university mailing list. Most of them ($N=16$) reported being somewhat or moderately familiar with VR technologies (Median = 2, IQR = 1, on a scale from 1 to 5). 
Regarding interacting with 2D artifacts in VR,
the most common 2D artifacts they interacted with were through web pages ($N=4$) and online searches in a browser ($N=3$). 
The study took around 70 to 90 minutes per session, and each participant was compensated with \$20 USD.

\subsection{Procedure, Apparatus, and Study Setup}
Nine pairs of participants ($N=18$) were invited to a room partitioned into two sections. The study implementation and apparatus can be referred to \autoref{sec:implementation}.
Upon completion of the consent form, participants were instructed to start with a warm-up task and then complete two tasks that cover different use case scenarios of Thing2Reality. \autoref{fig:Tasks} shows the scene configuration and the tasks.
After finishing each task, participants were asked to fill out a task-related survey. At the end of the study, we conducted semi-structured interviews with participants to gain more insights on the benefits, limitations, experiences with 2D and 3D representation of objects, and potential applications of Thing2Reality. We conducted statistical analysis on the survey answers, and distilled insights together with participants' subjective feedback.
All the sessions were video-recorded as first-person and third-person views for the post-study analysis.

\subsubsection{Walkthrough \& Avatar Decoration Task (30 min, \autoref{fig:Tasks}a-b)} 
The tutorial procedure was similar to the controlled study. In addition,
the two participants were instructed to search for personalized things such as hats, bags, umbrellas, and coffee mugs online, convert them into 3D objects, and use them to decorate their avatars. 
We chose this to familiarize the participants with how basic images can be used to generate multi-views and 3D Gaussians. After demonstrating with several examples, we encouraged the participants to conduct their own searches for objects to convert and use as decoration for their avatars.

\subsubsection{Furniture Arrangement Task (10 min, \autoref{fig:Tasks}c)}
The two participants collaborated in a shared virtual room and searched online platforms for desired images of furniture. These images were then converted into 3D objects and placed within the virtual environment. Through verbal communication and discussion, participants were required to collaboratively decide on the selection and placement of furniture items, and reach consensus.
This task has been used by prior work~\cite{hindmarsh2000object} in understanding  object-focused interaction and communication in virtual environment.
 
\subsubsection{Ideation and Pitch Task (20 min, \autoref{fig:Tasks}d)}
The two participants engaged in a spontaneous discussion about toy design that mimics a workspace scenario. There were three phases of the workspace task: \textit{selection phase}, \textit{preparation phase}, and \textit{pitch phase}.
First, each of them searched a toy from Google Images and decided which toy to pitch. They then prepared an elevator pitch about how to play with the toy. We provided several examples to the participants as reference. After the preparation, they performed a collaborative 2-minute pitch. During both preparation and presentation phases, they could use the whiteboard and 3D objects as assistance.

\subsection{Results}

The quantitative results are reported in \autoref{explore-quant}.
Overall, participants liked that they can spontaneously turn any images into 3D, which is more helpful than dedicated models, \eg, P10 articulated - ``\textit{...using online images was helpful rather than search for dedicated models.}'' P18 commented -  ``\textit{I like its ability to take 2D things on the webpage and convert them to 3D, I think it increases my options to find objects a lot more.}'' Furthermore, two participants mentioned that the 2D cropping and 3D object generation helped preserve the privacy compared to sharing the whole screen - P9 mentioned - ``\textit{you can see what your partner is talking about even if you can't see the browser. it's good privacy}''

\subsubsection{\textbf{Preferences and Trade-offs Between 2D Multiviews and 3D Gaussian Representations}}
Most participants ($n=16$) preferred 3D objects over the 2D Pie Menu ($n=2$) with orthogonal views (\autoref{fig:dataVisBar.png}). However, 2D orthogonal views on the Pie Menu were perceived as being easier to follow, having better image quality, and being more familiar to interact with. The ``unfolded effect'' of seeing different perspectives together in the Pie Menu was seen as useful and efficient by over half of the participants, as it required less manipulation than 3D objects, \eg, P2 commented, ``\textit{2D menu gave me a quick preview of views and took less effort than the 3D manipulation}''.

We probed participants' perceived comprehension and mental efforts for two phases: \textbf{(P1)} the phase of their self-cognitive and examination process when orthogonal views and 3D Gaussians were generated from their selected image (as the objects are spontaneously generated rather than pre-prepared); \textbf{(P2)} the phase when they use 2D snapshots and 3D Gaussians to communicate with their partners.

\paragraph{\textbf{P1}: Personal Comprehension Phase}
The participants expressed divided opinions over the mental efforts required for comprehension using the Pie Menu and 3D objects. While most participants found 3D objects requiring less mental effect, a few participants felt that 2D orthogonal views made it easier to understand different perspectives. For example, P10 expressed, ``\textit{Easiest is the 2D menu with orthogonal views. Less easy is 2D screen shots (snapshots) since I have to select the right view and move it around. Hardest is 3D view since moving changes how the item actually looks.}'' In contrast, P17 stated, ``\textit{3D object is quite easy to understand, the 2D snapshots and orthogonal views take some effort to use, and I don't think they are as helpful as 3D objects.}'' The remaining participants found both options easy to understand, with P4 expressing, ``\textit{2D is same as using computer, and 3D is closer to reality.}''

For personal comprehension of the generated objects, 10 participants thought that 3D objects were more intuitive and straightforward, while 4 preferred 2D snapshots continuously taken from any angle, and 2 preferred orthogonal views displayed on the Pie Menu (\autoref{fig:dataVisBar.png}: Right-Top). Participants who liked the orthogonal views on the Pie Menu mentioned that it was more understandable as they could clearly observe four sides of the object: \textit{``see the top, left, right, and bottom makes it easier to retrieve.''} Some preferred 2D snapshots since they felt it was familiar and easier to select images from different sides, \eg, ``\textit{I think it's easy to select 2D snapshots from different angles''} (P14). The majority of participants who preferred 3D objects appreciated the authenticity ($n=4$), interactivity ($n=4$), and comprehensive perspectives ($n=3$) that 3D representations offer, which largely overlap with the general feedback that they provided. 
For example, while completing the tasks, two participants specifically mentioned the desire to hold 3D objects in their hands for better clarification and presentation, \eg, P1 shared, \textit{``This was demonstrated more fully when conducting a workspace experiment. When I need to present a toy, holding it in my hand is more sales-oriented''}, and P5 discussed, \textit{``it's easier to argue when you have the object in hand''}.

\paragraph{\textbf{P2:} Communication Phase} 
When it came to the collaboration and communication phase, most participants ($n=12$) preferred 3D objects as they facilitated intuitive explanations and hands-on experiences. However, a few participants preferred 2D snapshots ($n=2$) for collaboration, as they provide the same perspective to everyone without navigating around the object (\autoref{fig:dataVisBar.png}: Right-Bottom). P10 commented, ``\textit{The way things look from any angle is the same. You don't have to be standing in one specific spot to see what I am seeing.}'' Furthermore, 2D snapshots were seen as useful for professional uses like presentations, as they made it easier to explain particular parts of an object without manipulating the 3D view. One participant mentioned, ``\textit{it helped us provide the businessperson with multiple views of the object}'' (P6), and P2 added, ``\textit{when trying to explain a particular part of the object, I think the snapshot made it easier to do since we didn't have to turn the 3D object and show it to the audience's perspective}.''

On the other hand, the participants found that 3D objects facilitated communication and made it easier to explain things, providing hands-on experiences and a greater sense of realism. P1 mentioned, ``\textit{it is much more detailed that it has all perspectives, so it is easier to explain objects intuitively. I don't have to take multiple screenshots of all the perspectives to explain some parts of the object.}'' This indicates that while 2D multi-views provide a quick and effortless way to preview different angles, the manipulable 3D Gaussian enable more detailed and intuitive explanations of object features.

\subsubsection{\textbf{Observation of Actual Usage of 2D and 3D During Presentation Phase}}


During the study, we observed that participants not only used 3D objects for discussion but also frequently employed them as a proxy for creating 2D snapshots. This finding aligns with our design goal of enabling flexible bi-directional transitions in XR workspaces (\textbf{DG3}).

In the workspace task, participants generally preferred 3D objects for discussion during their pitch preparation. However, they often use 2D representations in their final pitch deliverables. 
Only three out of the nine pairs used 3D objects in their final pitches, mainly to demonstrate object motion, such as showcasing a toy car's movement. The 3 pairs switched between 3D object showcase to referencing 2D artifacts pinned to the whiteboard. 

P2 commented on the complementary roles of 3D and 2D representations in presentations, stating, ``\textit{For presentation, 3D objects are better for communicating the actions or interactions with objects, while 2D can support a better organization.}'' This highlights the importance of having the flexibility to choose the most appropriate representation based on the specific communication needs.
Two participants mentioned that incorporating both 3D and 2D elements in a presentation might increase the mental effort required from the presenter. However, most of them recognized that this combination could be a valuable addition for the audience, as it helps them better understand the content being presented.

In brief, by allowing users to seamlessly switch between 3D and 2D representations, Thing2Reality empowers them to adapt their communication style to the specific requirements of the task at hand, which leads to more effective collaboration and clearer presentations.
\section{Discussion}
Our findings demonstrate Thing2Reality's ability to enhance communication and collaboration by transforming 2D content into multiview renderings and interactive 3D models, which opens up new possibilities for engaging and immersive experiences.  Thing2Reality not only functions as a prototype, but also advances our understanding of how 2D-3D cohabitation and transformation enhances communication in XR meetings.

\subsection{Is 3D More Effective and Efficient for Communication?}
For the comparative study that compares 2D and 3D formats, participants generally preferred 3D objects generated by Thing2Reality over 2D representations, valuing their  authenticity, immersion, different perspectives, and interactivity. Furthermore, participants showcased how to use the object more intuitively in 3D condition than the 2D condition.

During the exploratory study that explores the use of both 2D and 3D objects, 2D orthogonal views and snapshots were perceived as easier to follow, offering better image quality, and greater familiarity. Some participants reported that 3D objects can potentially occlude another user's view, especially for face-to-face configurations, reflecting prior findings on 2D versus 3D spaces~\cite{hauber2006spatiality}, whereas 2D snapshots on the interactive whiteboard coerces a consistent perspective for all users. 

Furthermore, our findings suggest that the 2D Pie Menu with four orthogonal views effectively supports an unfolded spatial perception of 3D objects. 
Extending the Pie Menu to include additional views, such as those at 45\textdegree intervals for a total of eight perspectives, could potentially enhance spatial understanding but increase cognitive loads.

While 3D objects were generally preferred for their intuitiveness, realistic detail, and spatial interactivity, 2D representations excelled in quick perspective-switching, consistent viewpoints for collaboration, and static presentation formats. Our findings indicate that users valued access to both 2D and 3D representations, as each fulfills distinct purposes and use cases. This suggests that balancing 2D and 3D representations can optimal user experiences in object sharing and collaboration scenarios.

Furthermore, 2D artifacts proved particularly useful during the pitch phase of the workspace task. 
These findings highlight the potential for flexible bi-directional transitions among digital media forms in XR workspaces. By enabling users to seamlessly switch between 3D and 2D representations, Thing2Reality allows users to adapt their communication style to the specific requirements of the task, fostering more effective collaboration and clearer presentations.

\subsection{Expectation of Generated Objects for Physical vs. Digital Objects}
We explored two data sources including physical object capturing and digital search as input techniques for creating 2D or 3D artifacts. Participants found both methods intuitive and appreciated the flexibility to source objects from online searches and their physical surroundings. The ability to combine physical and digital objects enhanced engagement and communication effectiveness.

While physical object capturing offered realistic references, participants desired more detail and accuracy. Converting digital objects (from digital search) into 3D provided a wide variety of items since most items only have one perspective/view point available only, but unpredictable sizing was a challenge, since participants do not have physical copies in their hand. These findings highlight the need for accurate size estimation in digital generation and improved detail retention in physical capture.

For physical object scanning, compared to existing methods like NeRF and traditional 3D reconstruction~\cite{mose2024}, our approach using 3D Gaussian splatting and multi-view diffusion models, while lacking in some accuracy (since other three view points were generated), demonstrated better support for quick idea demonstration with real-world items. It also showed superior performance and efficiency in generating perspectives of large objects (\eg, shelves) that are challenging to capture (when certain sides are occluded, or need to capture at a distance to get the whole picture) with conventional scanning methods.

The flexibility of input sources proved crucial, allowing users to incorporate both physical and digital objects into conversations seamlessly. However, balancing user expectations for output quality, particularly in size and detail accuracy for physical objects in their surrounding spaces, remains a key challenge for future development.

\subsection{The Potential of Generative AI in Enhancing Object-Centric XR Communication Experiences}
The study highlights the potential of Generative AI in creating authentic, immersive, and interactive object sharing experiences. Participants appreciated the ability to spontaneously turn 2D images into 3D objects, which expanded their options for finding and sharing objects. Generated 3D objects allowed users to better understand the functionality and user experience of the shared objects, as if they were real. This finding suggests that 
Generative AI could significantly enhance object-sharing experiences by offering more engaging and intuitive ways to communicate and collaborate around objects.

Some participants noted that the cropping and 3D object generation features of the system helped preserve privacy compared to sharing the whole screen. This highlights the potential of Thing2Reality in addressing privacy concerns in collaborative environments. However, participants also mentioned unexpected effects from Thing2Reality, particularly the GenAI models, such as the addition of unclear or unexpected elements to the generated 3D objects. This points to challenges related to the accuracy and interpretability of generated content.

\section{Limitations and Future Work}

\subsubsection*{Visualization of Abstract Ideas}
Thing2Reality has limitations in visualizing abstract ideas and concepts that lacks clear physical representations. This can pose challenges and lead to potential inaccuracies in professional communication scenarios, such as a delicate medical diagnosis~\cite{10.1145/3025453.3025566}. Future work could explore ways to represent abstract concepts more effectively, perhaps through the use of metaphors or symbolic representations.

\subsubsection*{Fidelity of Objects}

The fidelity of generated 3D Gaussians can be improved by increasing the input density. However, that will result in more time spent on the rendering for the current state~\cite{xu2024grm}.
Future work could investigate the integration of these methods with Thing2Reality to improve the visual quality of the generated objects while maintaining interactive performance.

Prior work explored fabricating on-screen virtual objects~\cite{10.1145/3450741.3465239} for game settings, which also shows the potential of this system. Once the generated 3D Gaussians are turned into mesh, they can potentially be converted into real, tangible, 3D-printed objects in the physical reality. This opens up exciting possibilities for rapid prototyping and physical visualization, but is not under the scope of this work.

\subsubsection*{Object-Level versus Scene-Level 3D Gaussian}

While our system explored 3D object-level interactions, Gaussian splatting can also be used to generate 3D scenes~\cite{kerbl20233d}. This can be extended to the exploration of the world of miniatures~\cite{10.1145/3411764.3445098,10.1145/223904.223938} utilizing the current state of the art. However, this would require more exploration into the manipulation and potential pros and cons of interacting with 3D scenes compared to individual objects. Future work could investigate the scalability of Thing2Reality to handle larger and more complex 3D scenes while maintaining usability and performance.

\subsubsection*{Automation of Human Input}
In this work, from an HCI perspective, we want to explore a way to use images (or objects of interest) as a \textit{middle ground} to bind a person's input (text, sketches, and digital search) and their intention to communicate with various and flexibly convertable data representations spontaneously shown to others (between 2D and 3D) to facilitate communication with other people. 
However, the current system requires manual input of text, sketches, and images to generate 3D objects. While this allows for greater control and customization, it may limit the system's efficiency. Future work could explore ways to automate or streamline the input process, such as using computer vision techniques to automatically extract object information from images or videos. Additionally, the system could be enhanced to adaptively understand the conversation context and suggest relevant 3D objects or visualizations based on the ongoing discussion. This would require the development of advanced natural language processing and machine learning algorithms to analyze and interpret the conversation in real-time.

\section{Conclusion} 

We believe that XR communication has tremendous promise for co-presence and for bridging distances between humans, yet much focus today is on realistic rendering of avatars and remote participants. However, as XR systems mature and become increasingly realistic, it will also become increasingly important to support a similar level of spontaneity with objects and artifacts, as what people experience in real environments.

In this paper, we presented Thing2Reality, an XR communication system that allows users to instantly materialize ideas or physical objects and share them as interactive conditioned multiview renderings or 3D Gaussians for realistic 3D rendering. 
Our first study ($N=12$) underscored Thing2Reality's effectiveness in elevating discussions by converting both physical and digital items into interactive 3D Gaussians, offering a more dynamic and comprehensive experience, compared to traditional 2D representations.
We further report on findings from our exploratory user study ($N=18$) that shows how the ability to interact with, and manipulate, both 2D and 3D object representations has the potential to significantly enhance discussions. 

Thing2Reality is one of many necessary building blocks towards increasingly realistic co-presence in XR, and we hope that our work will inspire continued work in this exciting domain.

\bibliographystyle{ACM-Reference-Format}
\bibliography{Thing3D}
\appendix
\section{Appendix}

\subsection{Quantitative Results of Exploratory Study}
\label{explore-quant}
\begin{figure*}
    \centering
    \includegraphics[width=\linewidth]{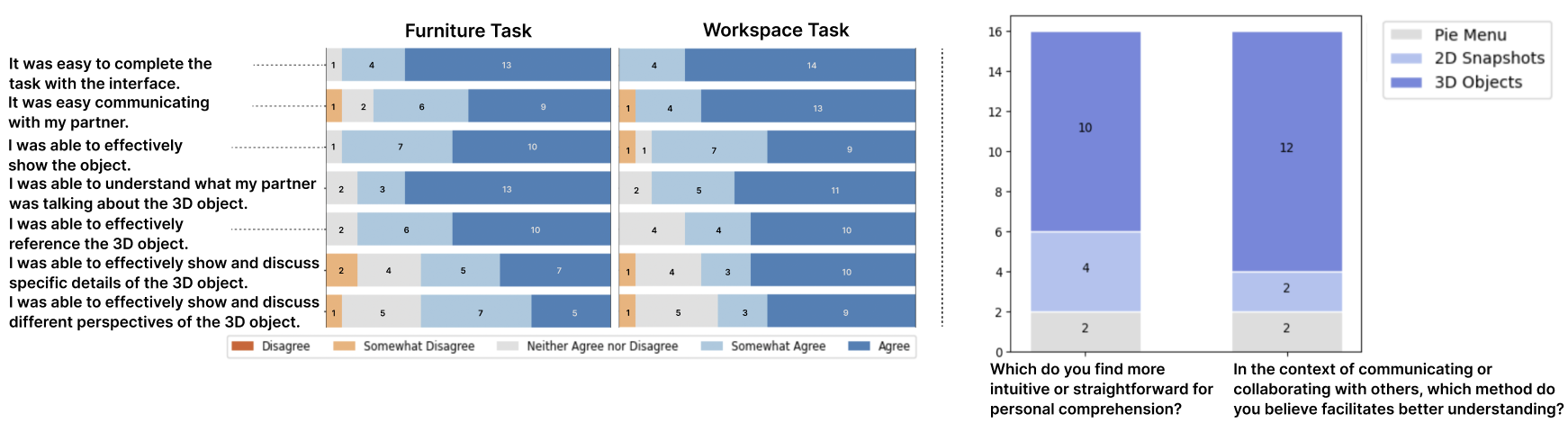}
    \caption{Survey Results. On the left, two stacked bar charts show the participants' responses in using Thing2Reality for the furniture and workspace tasks respectively. In both tasks, most participants reported Thing2Reality being easy to use and effective in communication and presentation of ideas through 3D objects. On the right, the stacked bar graph shows participants' preferences of 2D/3D representations when considering personal comprehension of the object and the ability to understand their collaborators. 3D objects were favored in both cases. }
    \label{fig:dataVisBar.png}
\end{figure*}
In both tasks (Furniture Arrangement and Workspace), the participants reported high system usability score: easy to communicate with the partner (Median = 5, IQR = 0.5; on a scale from 1 [``Disagree''] to 5 [``Agree'']), easy to complete the task with the interface (Median = 5, IQR = 1), and easy to use the interface (Median = 5, IQR = 1). They felt confident in both showing the 3D object effectively to their partners (Median = 5, IQR = 1) and understanding which part of the object their partners were referring to (Median = 5, IQR = 1). When comparing different object representations, the participants generally believed that they could deliver their ideas more clearly with 3D objects (Median = 5, IQR = 1) than the Pie Menu of multiple views (Median = 4, IQR = 2).

\end{document}